\documentclass[prd,superscriptaddress,notitlepage,showpacs,nofootinbib,longbibliography]{revtex4-1}

\usepackage{graphicx}
\usepackage[utf8]{inputenc} 
\usepackage{placeins}
\usepackage{amsmath}
\usepackage{amssymb}
\usepackage{float}
\usepackage{xcolor,color,colortbl}
\usepackage{comment}
\usepackage{multirow}
\usepackage{color}
\usepackage{soul}

\def\beq{\begin{equation}}
\def\eeq{\end{equation}}
\def\bea{\begin{eqnarray}}
\def\eea{\end{eqnarray}}

\usepackage{amsbsy}
\usepackage{bm}
\usepackage{color}
\usepackage{fancyhdr}
\usepackage[pdftex]{epsfig}
\usepackage[colorlinks=true,linkcolor=blue,citecolor=purple,urlcolor=magenta,filecolor=blue]{hyperref}
\usepackage{slashed}

\begin{document}

\bigskip

\vspace{2cm}

\title{$B$ meson anomalies within the triplet vector boson model to the light of recent measurements from LHCb}
\vskip 6ex

\author{J. M. Cabarcas}
\email{josecabarcas@usta.edu.co}
\affiliation{Universidad Santo Tom\'as, Colombia}
\author{J. H. Mu\~{n}oz}
\email{jhmunoz@ut.edu.co}
\affiliation{Departamento de F\'{i}sica, Universidad del Tolima, C\'{o}digo Postal 730006299, Ibagu\'{e}, Colombia}
\author{N\'{e}stor Quintero}
\email{nestor.quintero01@usc.edu.co}
\affiliation{Facultad de Ciencias B\'{a}sicas, Universidad Santiago de Cali, Campus Pampalinda, Calle 5 No. 62-00, C\'{o}digo Postal 76001, Santiago de Cali, Colombia}
\author{Eduardo Rojas}
\email{eduro4000@gmail.com}
\affiliation{Departamento de Física, Universidad de Nari\~no, A.A. 1175, San Juan de Pasto, Colombia}

\bigskip
  
\begin{abstract}
The triplet vector boson (TVB) is a simplified new physics model involving massive vector bosons transforming as a weak triplet vector. Such a model has been proposed as a combined explanation of the anomalous $b \to s\mu^+\mu^-$ and $b \to c \tau\bar{\nu}_\tau$ data (the so-called $B$ meson anomalies). In this work, we carry out an updated view of the TVB model by incorporating the most recent 2022 and 2023 LHCb measurements on the lepton flavor universality ratios $R(D^{(\ast)}) = {\rm BR}(B \to D^{(\ast)}\tau \bar{\nu}_\tau)/{\rm BR}(B \to D^{(\ast)}\ell^\prime \bar{\nu}_{\ell^\prime})$, $R(\Lambda_c) = {\rm BR}(\Lambda_b \to \Lambda_c\tau\bar{\nu}_\tau)/{\rm BR}(\Lambda_b \to \Lambda_c\mu\bar{\nu}_\mu)$, and $R_{K^{(\ast)}} = {\rm BR}(B \to K^{(\ast)}\mu^+\mu^-)/{\rm BR}(B \to K^{(\ast)}e^+e^-)$.
We perform a global fit to explore the allowed parameter space by the new data and all relevant low-energy flavor observables. Our results are confronted with the recent high-mass dilepton searches at the Large Hadron Collider (LHC). 
We find that for a heavy TVB mass of 1 TeV a common explanation of the $B$ meson anomalies is possible for all data with the recent LHCb measurements on $R(D^{(\ast)})$, in consistency with LHC constraints.
However, this framework is in strong tension with LHC bounds when one considers all data along with the world average values (BABAR, Belle, and LHCb) on $R(D^{(\ast)})$. Future measurements will be required in order to clarify such a situation.
In the end, the implications of our phenomenological analysis of the TVB model to some known flavor parametrizations are also discussed.
\end{abstract}

\maketitle


\section{Introduction}

In the last ten years, approximately, the high-energy physics community has been a witness of discrepancies between experimental measurements and the Standard Model (SM) calculations in several observables involving $b \to s\mu^+\mu^-$ (neutral-current) and $b \to c \tau\bar{\nu}_\tau$ (charged-current) transitions, which provide an important test of lepton flavor universality (LFU). Such inconsistencies indicate strong signals of LFU violation (for very recent interesting reviews, see Refs.~\cite{London:2021lfn,Albrecht:2021tul,Bifani:2018zmi}). For the neutral-current $b \to s\mu^+\mu^-$ transition, the ratio of semileptonic decay channels,
\begin{eqnarray}
R_{K^{(\ast)}} = \frac{{\rm BR}(B \to K^{(\ast)}\mu^+\mu^-)}{{\rm BR}(B \to K^{(\ast)}e^+e^-)},
\end{eqnarray}

\noindent provides a test of $\mu/e$ LFU for different dilepton mass-squared range $q^2$ ($q^2$ bins). From 2014 to 2021, the LHCb experiment reported the existence of discrepancies between the SM predictions and the experimental measurements (low and central $q^2$ bins) of $R_K$, $R_{K^\ast}$, $R_{K_S}$, and $R_{K^{\ast +}}$~\cite{Aaij:2014ora,Aaij:2019wad,Aaij:2017vbb,Aaij:2021vac,LHCb:2021lvy}, hinting toward LFU violation in the $\mu/e$ sector. However, at the end of 2022, an improved LHCb analysis of the ratios $R_{K^{(\ast)}}$, namely~\cite{LHCb:2022qnv,LHCb:2022zom}
\begin{eqnarray}
R_{K} &=&
\begin{cases}
 0.994^{+0.090 + 0.029}_{-0.082 - 0.027},  \ \  q^2 \in [0.1,1.1] \ {\rm GeV}^2, \\
 0.949^{+0.042 + 0.022}_{-0.041 - 0.022},  \ \ q^2 \in [1.1,6.0] \ {\rm GeV}^2,
\end{cases}
\end{eqnarray}
and
\begin{eqnarray}
R_{K^\ast} &=&
\begin{cases}
  0.927^{+0.093 + 0.036}_{-0.087 - 0.035},  \ \  q^2 \in [0.1,1.1] \ {\rm GeV}^2, \\
 1.027^{+0.072 + 0.027}_{-0.068 - 0.026},  \ \ q^2 \in [1.1,6.0] \ {\rm GeV}^2,
\end{cases}
\end{eqnarray}

\noindent now shows a good agreement with the SM~\cite{LHCb:2022qnv,LHCb:2022zom}. In addition, the CMS experiment has presented a new measurement of the branching ratio of $B_s \to \mu^+\mu^-$ more consistent with the SM~\cite{CMS:2022mgd}. 
Despite that the tension on $R_{K^{(\ast)}}$ ratios and ${\rm BR}(B_s \to \mu^+\mu^-)$ has now disappeared, there are still some discrepancies in the measurements of additional $b \to s\mu^+\mu^-$ observables, such as angular observables and differential branching fractions related with $B \to K^\ast  \mu^+\mu^-$ and $B_s \to  \phi  \mu^+\mu^-$ decays~\cite{Aaij:2013qta,Aaij:2015oid,
Aaij:2020nrf,Aaij:2013aln,Aaij:2015esa,Aaij:2020ruw}. 
Within a model-independent effective Hamiltonian approach and under the hypothesis that New Physics (NP) couples selectively to the muons, different scenarios with NP operators (dimension-six) have been surveyed in the literature~\cite{Aebischer:2019mlg,Altmannshofer:2021qrr, Alguero:2021anc,Alguero:2019ptt,Geng:2021nhg,Hurth:2021nsi,Angelescu:2021lln,Carvunis:2021jga,London:2021lfn,Greljo:2022jac,Alguero:2023jeh}.
The most recent global fit analysis~\cite{Greljo:2022jac,Alguero:2023jeh} taking into account updated $b \to s \mu^+\mu^-$ data (including $R_{K^{(\ast)}}$ by LHCb~\cite{LHCb:2022qnv,LHCb:2022zom} and ${\rm BR}(B_s \to \mu^+\mu^-)$ by CMS~\cite{CMS:2022mgd}), showed that the Wilson coefficient (WC) solution $C^{bs\mu\mu}_{9} = - C^{bs\mu\mu}_{10}$, related with the operators $(\bar{s}P_{L} \gamma_\alpha b) (\bar{\mu}\gamma^\alpha \mu)$ and $(\bar{s}P_{L} \gamma_\alpha b) (\bar{\mu}\gamma^\alpha\gamma_5 \mu)$, is still a viable solution to describe the data.\medskip

On the other hand, the experimental measurements collected by the BABAR, Belle, and LHCb experiments on different charged-current $b \to c \tau\bar{\nu}_\tau$ observables, indicate the existence of
disagreement with respect to the SM predictions~\cite{Lees:2012xj,Lees:2013uzd,Huschle:2015rga,Sato:2016svk,Hirose:2017vbz,Aaij:2015yra,
Aaij:2017deq,Aaij:2017uff,Belle:2019rba,Hirose:2017dxl,Hirose:2016wfn,Abdesselam:2019wbt,
HFLAV:2022pwe,LHCb2022,LHCb:2023zxo,LHCb2023,HFLAVsummer,Aaij:2017tyk} (see Table~\ref{Table:1} for a summary). Regarding the ratios of semileptonic $B$ meson decays, 
\begin{equation} \label{RD}
R(D^{(\ast)}) = \dfrac{{\rm BR}(B \to D^{(\ast)}\tau \bar{\nu}_\tau)}{{\rm BR}(B \to D^{(\ast)}\ell^\prime \bar{\nu}_{\ell^\prime})},
\end{equation}

\noindent with $\ell^\prime = e \ {\rm or} \ \mu$ (the so-called \textit{$R(D^{(\ast)})$ anomalies}),  the LHCb has presented, very recently, the first combined measurement using Run 1 data (3 fb$^{-1}$) with muonic $\tau$ decay reconstruction~\cite{LHCb2022,LHCb:2023zxo},
\begin{eqnarray}
R(D)_{\rm LHCb22} &=&  0.441 \pm 0.060 \pm 0.066, \\
R(D^{\ast})_{\rm LHCb22} &=& 0.281 \pm 0.018 \pm 0.024  ,
\end{eqnarray}

\noindent which show a tension of $1.9\sigma$ with the SM predictions. Additionally, the LHCb also reported a preliminary measurement of $R(D^{\ast})$ using partial Run 2 data (2 fb$^{-1}$), where the $\tau$ is hadronically reconstructed~\cite{LHCb2023}. When combined with Run 1, the result is~\cite{LHCb2023} 
\begin{equation}
R(D^{\ast})_{\rm LHCb23} = 0.257 \pm 0.012 \pm 0.018, 
\end{equation}

\noindent that is compatible with SM at the $\sim 1\sigma$ level. Incorporating these new LHCb results, the preliminary 
world average values reported by the Heavy Flavor Averaging Group (HFLAV) are~\cite{HFLAVsummer}
\begin{eqnarray}
R(D)_{\rm HFLAV23} &=&  0.356 \pm 0.029, \\
R(D^{\ast})_{\rm HFLAV23} &=& 0.284 \pm 0.013  ,
\end{eqnarray}

\noindent that now exceed the SM by $3.2\sigma$. Moreover, the LHCb measurement of the ratio $R(J/\psi) = {\rm BR}(B_c \to J/\psi \tau \bar{\nu}_\tau)/{\rm BR}(B_c \to J/\psi\mu \bar{\nu}_{\mu})$~\cite{Aaij:2017tyk} also shows tension ($\sim 2\sigma$) with regard to the SM prediction~\cite{Harrison:2020nrv}. Additional hints of LFU violation in the $b \to c \tau\bar{\nu}_\tau$ transition have been obtained in the Belle measurements of the $\tau$ lepton polarization $P_\tau(D^\ast)$~\cite{Hirose:2017dxl,Hirose:2016wfn} and the longitudinal polarization of the $D^*$ meson $F_L(D^\ast)$~\cite{Abdesselam:2019wbt} related with the channel $\bar{B} \to D^\ast \tau \bar{\nu}_\tau$, which also exhibit a deviation from the SM values~\cite{Iguro:2022yzr}. 
 The tauonic channel $B_c \to J/\psi \tau \bar{\nu}_\tau$ has not been measured yet, but indirect constraints on its branching ratio have been imposed $<30\%$~\cite{Alonso:2016oyd} and $<10 \%$~\cite{Akeroyd:2017mhr}.
In Table~\ref{Table:1} we summarize the current experimental measurements and their corresponding SM predictions. We also collect in Table~\ref{Table:1} the experimental and theoretical values of the ratio of inclusive decays $R(X_c) \equiv {\rm BR}(B\to X_c\tau \bar{\nu}_\tau)/{\rm BR}(B\to X_c\mu \bar{\nu}_\mu)$, which is generated via the same $b \to c \tau\bar{\nu}_\tau$ transition~\cite{Kamali:2018bdp}.
The SM estimation on $R(X_c)$ is based on the $1S$ mass scheme and includes nonperturbative corrections of the order $\mathcal{O}(1/m_b^3)$, while the NP effects took into account the subleading $\mathcal{O}(1/m_b)$ corrections~\cite{Kamali:2018bdp}. 
The $R(D^{(\ast)})$ anomalies still exhibit the largest deviation. The other $b \to c \tau\bar{\nu}_\tau$ observables also show tension (moderate) with the data, although, some of them have large experimental uncertainties (such as $R(J/\psi)$ and $P_\tau(D^\ast)$). While the ratio $R(X_c) $ is in excellent agreement with the SM. \medskip

In addition, the LHCb Collaboration has recently released the first measurement of the ratio of semileptonic $\Lambda_b$ baryon decays, namely~\cite{LHCb:2022piu} 
\begin{equation} \label{R_Lambda_c}
R(\Lambda_c) \equiv \dfrac{{\rm BR}(\Lambda_b^0 \to \Lambda_c^+ \tau^-\bar{\nu}_\tau)}{{\rm BR}(\Lambda_b^0 \to \Lambda_c^+ \mu^-\bar{\nu}_\mu)} = 0.242  \pm 0.076,
\end{equation}

\noindent in agreement at the $\sim 1.2\sigma$ level with the most recent SM calculation, $R(\Lambda_c)_{\rm SM} = 0.324 \pm 0.004$~\cite{Bernlochner:2018bfn}. In Eq.~\eqref{R_Lambda_c} we have added in quadrature the statistical and systematic uncertainties, and the external branching ratio uncertainty from the channel $\Lambda_b^0 \to \Lambda_c^+ \mu^-\bar{\nu}_\mu)$~\cite{LHCb:2022piu}. It is interesting to highlight that this new measurement is below the SM value, pointing to an opposite direction than the current $b \to c \tau\bar{\nu}_\tau$ data (see Table~\ref{Table:1}). Nevertheless, in order to provide an overall picture, all the anomalous $b \to c \tau\bar{\nu}_\tau$ data must be taken into account. To the best of our knowledge, the impact of the new LHCb measurement on $R(\Lambda_c)$ has been recently studied from a model-independent way (effective field theory approach)~\cite{Fedele:2022iib} and in the singlet vector leptoquark model~\cite{Garcia-Duque:2022tti}.   \medskip

\begin{table}[!t]
\centering
\renewcommand{\arraystretch}{1.4}
\renewcommand{\arrayrulewidth}{0.8pt}
\caption{\small Experimental status and SM predictions on observables related to the charged-current transitions $b \to c \ell \bar{\nu}_\ell$ ($\ell = \mu, \tau$).}
\begin{tabular}{cccc}
\hline
Transition & Observable & Expt. measurement & SM prediction \\
\hline
$b \to c \tau\bar{\nu}_\tau$ & $R(D)$ & $0.441 \pm 0.060 \pm 0.066$ (LHCb22)~\cite{LHCb2022,LHCb:2023zxo} & 0.298 $\pm$ 0.004~\cite{HFLAVsummer} \\
 &  & $0.356 \pm 0.029$ (HFLAV)~\cite{HFLAVsummer} &  \\
 & $R(D^\ast)$ & $0.281 \pm 0.018 \pm 0.024$ (LHCb22)~\cite{LHCb2022,LHCb:2023zxo} & 0.254 $\pm$ 0.005~\cite{HFLAVsummer} \\
  &  & $0.257 \pm 0.012 \pm 0.018$ (LHCb23)~\cite{LHCb2023} & \\
  &  & $0.284 \pm 0.013$ (HFLAV)~\cite{HFLAVsummer} &  \\
 & $R(J/\psi)$ & $0.71 \pm 0.17 \pm 0.18$~\cite{Aaij:2017tyk} & 0.2582 $\pm$ 0.0038~\cite{Harrison:2020nrv}  \\
& $P_\tau(D^\ast)$ & $- 0.38 \pm 0.51 ^{+0.21}_{-0.16}$~\cite{Hirose:2017dxl,Hirose:2016wfn} &  $-0.497 \pm 0.007$~\cite{Iguro:2022yzr} \\
& $F_L(D^\ast)$ & $0.60 \pm 0.08 \pm 0.035$~\cite{Abdesselam:2019wbt} & $0.464 \pm 0.003$~\cite{Iguro:2022yzr}  \\
& $R(X_c)$ & 0.223 $\pm$ 0.030~\cite{Kamali:2018bdp} & 0.216 $\pm$ 0.003~\cite{Kamali:2018bdp} \\
& ${\rm BR}(B_c^- \to \tau^- \bar{\nu}_{\tau})$ & $<10 \%$~\cite{Akeroyd:2017mhr}, $<30\%$~\cite{Alonso:2016oyd}  & $(2.16 \pm 0.16 )\%$~\cite{Gomez:2019xfw} \\
\hline 
$b \to c \mu\bar{\nu}_\mu$ & $R_D^{\mu/e}$ & $0.995 \pm 0.022 \pm 0.039$~\cite{Glattauer:2015teq} & $0.9960\pm 0.0002$~\cite{Becirevic:2020rzi} \\
                           & $R_{D^\ast}^{\mu/e}$ & $0.961 \pm 0.050$~\cite{Belle:2017rcc} & $0.9974\pm 0.0001$~\cite{Bobeth:2021lya} \\
\hline
\end{tabular} \label{Table:1}
\end{table}

Although the $b \to c \tau\bar{\nu}_\tau$ data is suggesting stronger signals of LFU violation than $b \to s \mu^+\mu^-$ one, a combined explanation of the current data is still desirable. This simultaneous explanation can be generated by different tree-level heavy mediators with adequate couplings, for example, charged scalar bosons, extra gauge bosons or leptoquarks (scalar and vector). For an extensive list of literature, see the theoretical status report presented in Ref.~\cite{London:2021lfn}. In this work,  we will pay particular attention to the common explanation provides by the so-called Triplet Vector Boson (TVB) model~
\cite{Calibbi:2015kma,Bhattacharya:2014wla,Greljo:2015mma,Faroughy:2016osc,Buttazzo:2017ixm,Bhattacharya:2016mcc,
Kumar:2018kmr,Guadagnoli:2018ojc,Boucenna:2016wpr,Boucenna:2016qad}~\footnote{Let us notice that in a recent work~ \cite{Capdevila:2020rrl}, the TVB model was implemented as an explanation to the Cabibbo angle anomaly and $b \to s \ell^+\ell^-$ data.}, in which the SM is extended by including a color-neutral real $SU(2)_L$ triplet of massive vectors $W^\prime$ and $Z^\prime$ that coupled predominantly to left-handed (LH) fermions from the second- and third-generations~\cite{Calibbi:2015kma,Bhattacharya:2014wla,Greljo:2015mma,Faroughy:2016osc,Buttazzo:2017ixm,Bhattacharya:2016mcc,
Kumar:2018kmr,Guadagnoli:2018ojc,Boucenna:2016wpr,Boucenna:2016qad}. The neutral boson $Z^\prime$ is responsible for the $b \to s \mu^+\mu^-$ data, while the charged boson $W^\prime$  generates the $b \to c \tau\bar{\nu}_\tau$ one. 
We adopt a phenomenological approach of the TVB model based on the minimal setup of couplings between the new gauge bosons $Z^\prime, W^\prime$  and LH fermions of the SM, without specifying the complete UV model. 
We present an updated analysis of TVB model (parametric space) by including the new 2022 and 2023 LHCb data on $R_{K^{(\ast)}}$, $R(D^{(\ast)})$, and $R(\Lambda_c)$. We also incorporate in our study all relevant flavor observables that are also affected by this NP model, such as $B_s - \bar{B}_s$ mixing, neutrino trident production, LFV decays ($B \to K^{(\ast)} \mu^\pm \tau^\mp$, $B_s \to \mu^\pm \tau^\mp$, $\tau \to \mu\phi$, $\Upsilon(nS) \to \mu^\pm \tau^\mp$), rare $B$ decays ($B \to K^{(\ast)} \nu\bar{\nu}, B \to K \tau^+ \tau^-, B_s \to \tau^+ \tau^-$), and bottomonium LFU ratios. Furthermore, we study the consistency of the allowed TVB parameter space with the Large Hadron Collider~(LHC) bounds from searches of high-mass dilepton resonances at the ATLAS experiment. \medskip
 
Even though our focus will be phenomenological, regarding the ultra violet (UV) complete realization for the TVB model, the extension of the SM must allow for Lepton Flavor Non Universal (LFNU) couplings to the extra gauge bosons and LFV. In this direction in Ref.~\cite{Boucenna:2016qad} there is a proposal in which an extra $SU(2)$ gauge group is added and where extra scalars, new vector-like fermions and some non trivial transformations under the SM group are included. It is clear, that the couplings of fermions to the extra gauge bosons of the particular UV realization, will have model-dependent consequences that might relate different terms between them; however, since we make emphasis that our approach is phenomenological, we will start from the most general lagrangian for the TVB model as possible, and we will make comparisons to other approaches presented in Refs. \cite{Calibbi:2015kma,Greljo:2015mma,Buttazzo:2017ixm,Bhattacharya:2016mcc} where the new physics is coupled predominantly to the second and third generation of left handed quarks and leptons, ensuring LFNU and LFV through different mechanisms. Restrict our results to a particular UV-model is out of our target.
\medskip

This paper is structured as follows: in Sec.~\ref{TVB} we discuss the main aspects of the TVB model to accommodate the $B$ meson anomalies. As a next step in Sec.~\ref{Obs}, we consider the most relevant flavor observables and present the TVB model contributions to them. The LHC bounds are also studied. We then perform our phenomenological analysis of the allowed parametric space in Sec.~\ref{analysis} and our conclusions are presented in Sec.~\ref{Conclusion}.

\section{The Triplet Vector boson model} \label{TVB}
In general, flavor anomalies have been boarded into the current literature as a motivation to build innovative models and to test well established New Physics (NP) models. In this section, we focus in the previously mentioned Triplet Vector Boson (TVB) model \cite{Bhattacharya:2014wla,Calibbi:2015kma,Kumar:2018kmr,Faroughy:2016osc,Greljo:2015mma,Bhattacharya:2016mcc,Buttazzo:2017ixm,Guadagnoli:2018ojc,Boucenna:2016wpr,Boucenna:2016qad}  as a possible explanations of these anomalies, that might accommodate the observed flavor experimental results.   
 One significant feature of this model, is the inclusion of extra SM-like vector bosons with non-zero couplings to the SM fermions, that allow us to include additional interactions.\medskip

In the fermion mass basis, the most general lagrangian describing the dynamics of the fields can be written as
\begin{eqnarray}\label{inicial}
\Delta {\cal L}_V = g_{ij}^q(\bar{\Psi}_{iL}^Q\gamma^\mu\sigma^I \Psi^Q_{jL})V^I_\mu +g_{ij}^\ell(\bar{\Psi}_{iL}^\ell\gamma^\mu\sigma^I \Psi^\ell_{jL})V^I_\mu
\end{eqnarray}
 where, $V_\mu$ stands for the extra or new vector bosons  that transform as (1, 3, 0) under the $SU(3)_C\otimes SU(2)_L\otimes U(1)_Y$ gauge symmetry and must be redefined as $ W^{\prime\pm}, \, Z^\prime$. On the other side, SM fermions are arranged into the doublets $\Psi^Q_L$ and $\Psi^\ell_L$ given by 
\begin{eqnarray}\label{doublets}
\Psi^Q_{L} = \begin{pmatrix} V^\dag u_L\cr d_L \end{pmatrix}, \qquad \Psi^\ell_L=
 \begin{pmatrix} \nu_L\cr \ell_L \end{pmatrix}.
\end{eqnarray}
It is worth noticing here that in this particular model the CKM mixing matrix $V$  is applied on the up-type quarks.

In order to find the effective lagrangian for this model, the heavy degrees of freedom corresponding to vector bosons introduced above must be integrated out. Introducing the definition for the currents $J_{Q} = \bar{\Psi}_{iL}^Q\gamma^\mu\sigma^I \Psi^Q_{jL}$ and $J_{\ell} = \bar{\Psi}_{iL}^\ell\gamma^\mu\sigma^I \Psi^\ell_{jL}$, the effective lagrangian is therefore
 \begin{eqnarray}
 {\cal L}_{eff} &=& -\frac{(g_{ij}^q J_Q+ g_{ij}^\ell J_\ell)^2}{2M^2_{V}}\\
&=&  -\frac{(g_{ij}^q J_Q)^2}{2M^2_{V}}-\frac{g_{ij}^q g_{kl}^\ell J_Q  J_\ell}{M^2_{V}}-\frac{(g_{ij}^\ell J_\ell)^2}{2M^2_{V}}.
 \end{eqnarray}
The middle term of the right-hand side of the above equation corresponds to
\begin{eqnarray}\label{lageffexpand} 
\frac{g_{ij}^q g_{kl}^\ell J_Q  J_\ell}{M^2_{V}} &=& \frac{g_{ij}^q g_{kl}^\ell}{M_{V}^2}(\bar{\Psi}^Q_{iL}\gamma_\mu\sigma^I \Psi^Q_{jL})(\bar{\Psi}^\ell_{kL}\gamma^\mu\sigma^I \Psi^\ell_{lL})
\end{eqnarray}
Substituting  equation (\ref{doublets}) in the last expression, it leads us to
\begin{eqnarray}
\frac{g_{ij}^q g_{kl}^\ell J_Q  J_\ell}{M^2_{V}}  &=& 2\frac{ g_{kl}^\ell}{M_{V}^2}\left[(V g^d)_{ij}(\,\bar{u}_{iL}\gamma_\mu d_{jL})(\bar{\ell}_{k}\gamma^\mu\nu_{lL})+{\rm h.c.}\right]\nonumber\\
&&+\frac{ g_{kl}^\ell}{M_{V}^2}\left[(V g^dV^\dag)_{ij}(\bar{u}_{iL}\gamma_\mu u_{jL})(\bar{\nu}_{kL}\gamma^\mu \nu_{lL})+g_{ij}^d(\bar{d}_{iL}\gamma_\mu d_{jL})(\bar{\ell}_{kL}\gamma^\mu \ell_{lL})\right]\nonumber\\
&&-\frac{ g_{kl}^\ell}{M_{V}^2}\left[(V g^dV^\dag)_{ij}(\bar{u}_{iL}\gamma_\mu u_{jL})(\bar{\ell}_{kL}\gamma^\mu \ell_{lL})+g_{ij}^d(\bar{d}_{iL}\gamma_\mu d_{jL})(\bar{\nu}_{kL}\gamma^\mu \nu_{lL})\right],
\end{eqnarray}
in this expression, we can identify that the first term expresses an effective interaction of the SM fields that should be mediated by extra bosonic charged fields, while the remaining terms are mediated by an extra neutral bosonic field. These mediators are precisely the vector boson fields  ${W'}$ and ${Z'}$ introduced in this model and which masses can naively be considered to be (almost) degenerated which is required by electroweak precision data \cite{Faroughy:2016osc}. For simplicity, and without losing generality, we are going to consider that the couplings $g^{q,\ell}$ are real to avoid CP violation effects.  Additionally, it is important to notice that we can write compactly the couplings of quarks to the vector boson fields with an explicit dependence in the couplings of the down sector and also, keeping in mind that the CKM matrix couples into the doublets to up-type quarks and that we should restrict the significant contributions for the second and third families. For this purpose, we restrict the relevant couplings of the down sector to $g_{bb}$, $g_{ss}$ and $g_{sb} = g_{bs}$ while other terms remain zero.  This hypothesis that the couplings to the first generation of fermions (also in the leptonic sector) can be neglected has been widely accepted in the literature into the context of flavor anomaly explanations \cite{Calibbi:2015kma,Kumar:2018kmr,Greljo:2015mma,Faroughy:2016osc,Bhattacharya:2016mcc,Bhattacharya:2014wla,Buttazzo:2017ixm,Guadagnoli:2018ojc}. Lastly, the resultant compact form for the couplings of the quark sector to the $W^\prime$ that we obtained are
\begin{eqnarray}
 g_{\alpha b} &=& g_{bb}V_{\alpha b}+g_{sb}V_{\alpha s},\nonumber\\
 g_{\alpha s} &=& g_{ss}V_{\alpha s}+g_{sb}V_{\alpha b},
 \end{eqnarray}
where $\alpha$ stands for $u,c$ or $t$ quark flavors. The same procedure described above must be implemented for a compact form of the couplings of up-type quarks to the $Z^\prime$ boson. In this case we find two possibilities: one on flavor conserving interaction given by
\begin{eqnarray}
 g_{\alpha\alpha} &=& g_{bb}V_{\alpha b}^2+2g_{\alpha b}V_{\alpha s}V_{\alpha b}+g_{ss}V_{\alpha s}^2;
 \end{eqnarray}
 the other is related to flavor changing $Z^\prime$ couplings mediated by
 \begin{eqnarray}
 g_{\alpha \beta} =  g_{bb}V_{\beta b}V_{\alpha b}+g_{sb}V_{\beta s}V_{\alpha b}+g_{sb}V_{\beta b}V_{\alpha s}+ g_{ss}V_{\beta s}V_{\alpha s},
 \end{eqnarray}
 where $\alpha\neq \beta$ labels $u,c$ or $t$ quark flavors.

 To close this kind of parametrization, we mention that the terms of the r.h.s of equation (\ref{lageffexpand}) are responsible for and will be important to $4q$ and $4\ell$ interactions ruled by the lagrangian
 
\begin{eqnarray}
{\cal L}_{NP}^{4q,4\ell} = -\frac{g_{ij}^q g_{kl}^q}{2 M_{V}^2}(\bar{\Psi}^Q_{iL}\gamma_\mu\sigma^I \Psi^Q_{jL})(\bar{\Psi}^Q_{kL}\gamma^\mu\sigma^I \Psi^Q_{lL})-\frac{g_{ij}^\ell g_{kl}^\ell}{2 M_{V}^2}(\bar{\Psi}^\ell_{iL}\gamma_\mu\sigma^I \Psi^\ell_{jL})(\bar{\Psi}^\ell_{kL}\gamma^\mu\sigma^I \Psi^\ell_{lL})
\end{eqnarray}

\subsection{Other parametrizations}

In this subsection, we compare the previous parameterization explained above with others used in some representative references studied widely in the TVB model. 

In the TVB model presented in refs \cite{Calibbi:2015kma,Bhattacharya:2016mcc}, the mixing pattern for quarks is enriched by the inclusion of mixing matrices that will rotate the fields from the gauge basis to the mass basis and a projector ($X,Y$) that will ensure the dominance of the second and third families to explain anomalies. Particularly, the explicit form of these matrices for the down-type quarks and charged leptons and projectors are
\begin{eqnarray}
D=\begin{pmatrix}1 & 0 & 0\cr 0 &  \cos\theta_D & \sin\theta_D \cr 0 & -\sin\theta_D & \cos\theta_D\end{pmatrix},\qquad L=\begin{pmatrix}1 & 0 & 0\cr 0 &  \cos\theta_L & \sin\theta_L \cr 0 & -\sin\theta_L & \cos\theta_L\end{pmatrix},\qquad
X = Y = \begin{pmatrix}0 & 0 & 0\cr 0& 0 & 0\cr 0 & 0 & 1\end{pmatrix}.
\end{eqnarray}
These matrices will leave an explicit dependence of these mixing angles ($\theta_{D,L}$) into the couplings to the extra fields, which by the experimental results coming from different observables, can be constrained. The assumptions made in the introduction of these matrices were previously introduced in \cite{Calibbi:2015kma}, and we can establish the full equivalence between the notations of the angles by the relations $\theta_D = \alpha_{sb}$ and $\theta_L = \alpha_{\mu\tau}$. We also found that these couplings can be translated to the generic parameterization introduced at the beginning of  this section. For this purpose, as it was explained before, the couplings of all the quark sector will be dependent on the couplings of the down-type quarks, particularly in this kind of parameterization, we can illustrate the way that the couplings are obtained through the effective charged lagrangian that will be given as
\begin{eqnarray}
{\cal L}_{\rm eff}^{W^\prime} = 2\frac{ g_2^q g_2^\ell}{M_{V}^2}\left[(V\, D^\dag X \, D  )_{ij}(\,\bar{u}_{iL}\gamma_\mu d_{jL})( L^\dag Y \, L  )_{kl}(\bar{\ell}_{k}\gamma^\mu\nu_{lL})+{\rm h.c}\right];
\end{eqnarray}
thus, we obtain the equivalence 
\begin{eqnarray}\label{gtogquark}
g_{bb} &\to& g_2^q\cos^2\theta_D\nonumber\\
g_{sb} &\to& - g_2^q\sin\theta_D\cos\theta_D\nonumber\\ 
g_{ss} &\to&g_2^q\sin^2\theta_D,
\end{eqnarray}
and for the leptonic sector 
\begin{eqnarray}\label{gtoglepton}
g_{\tau\tau} &\to& g_2^\ell\cos^2\theta_L\nonumber\\ 
g_{\mu\tau}  &\to& - g_2^\ell\sin\theta_L\cos\theta_L\nonumber\\
g_{\mu\mu}   &\to&g_2^\ell\sin^2\theta_L. 
\end{eqnarray}

The comparison and equivalence among  parameterizations of different influential references  can be found in Tables~\ref{tab1},~\ref{tab2},~\ref{tab3} and~\ref{tab4}. \medskip

For our last comparison, we considered the parameterization given in Refs.~\cite{Greljo:2015mma,Buttazzo:2017ixm} where the couplings to the vector bosons have almost the same structure of the initial parameterization presented here, but its major difference consists in the dependence on flavor matrices denoted by the authors as $\lambda_{ij}^{(q,\ell)}$.  This incidence of the flavor structure into the model can be shown using the charged effective lagrangian as we did before
\begin{eqnarray}
{\cal L}_{\rm eff}^{W^\prime} = \frac{ g_q g_\ell}{2M_{V}^2}\left[(V\, \lambda  )_{ij}(\,\bar{u}_{iL}\gamma_\mu d_{jL})(\bar{\ell}_{k}\gamma^\mu\nu_{lL})+{\rm h.c}\right],
\end{eqnarray}
to obtain the desired dominance of couplings to the second and third families using the flavor matrices mentioned before, the $\lambda_{ij}$ belonging to the first family must be set to zero. Additionally, the values for $\lambda_{bb} =\lambda_{\tau\tau}=1$ in order to maximize its contribution.  However, as an illustration, we can make a complete relation of the implementation of the flavor matrices to the construction of couplings for the quark sector without any assumption in Tables~\ref{tab1},~\ref{tab2},~\ref{tab3} and~\ref{tab4}.

 \begin{table}[h!]
 \begin{center}
  \caption{\label{tab1}Couplings to $W^\prime$ boson in different parameterizations of the TVB model}
\begin{tabular}{ |c|c|c|c| } 
 \hline
 Coupling &Parameterization in \cite{Kumar:2018kmr} &  Parameterization in \cite{Calibbi:2015kma,Bhattacharya:2016mcc}& Parameterization in \cite{Greljo:2015mma,Buttazzo:2017ixm}\\ \hline
$g_{ub}^q$ & $g_{bb}V_{ub}+g_{sb}V_{us}$ &  $ g_2^q(V_{ub}\cos^2\theta_d-V_{us}\cos\theta_d\sin\theta_d)$ & $g_q(V_{ub}+V_{ud}\lambda_{db}+V_{us}\lambda_{sb})/\sqrt{2}$ \\ 
$g_{cb}^q$ & $g_{bb}V_{cb}+g_{sb}V_{cs}$ & $ g_2^q(V_{cb}\cos^2\theta_d-V_{cs}\cos\theta_d\sin\theta_d) $ & 
$g_q(V_{cb}+V_{cd}\lambda_{db}+V_{cs}\lambda_{sb})/\sqrt{2}$\\ 
$ g_{tb}^q$ & $g_{bb}V_{tb}+g_{sb}V_{ts}$ & $ g_2^q(V_{tb}\cos^2\theta_d-V_{ts}\cos\theta_d\sin\theta_d)$ & 
$g_q(V_{tb}+V_{ud}\lambda_{tb}+V_{us}\lambda_{sb})/\sqrt{2}$\\
$ g_{us}^q$ & $g_{ss}V_{us}+g_{sb}V_{ub}$ & $ g_2^q(V_{us}\sin^2\theta_d-V_{ub}\cos\theta_d\sin\theta_d)$ & 
$g_q(V_{ud}\lambda_{ds}+V_{ub}\lambda_{sb}+V_{us}\lambda_{ss})/\sqrt{2}$\\
$ g_{cs}^q$ & $g_{ss}V_{cs}+g_{sb}V_{ucb}$ & $ g_2^q(V_{cs}\sin^2\theta_d-V_{cb}\cos\theta_d\sin\theta_d)$ & $g_q(V_{cd}\lambda_{ds}+V_{cb}\lambda_{sb}+V_{cs}\lambda_{ss})/\sqrt{2}$\\
$g_{ts}^q$ & $g_{ss}V_{ts}+g_{sb}V_{tb}$ & $g_2^q(V_{ts}\sin^2\theta_d-V_{tb}\cos\theta_d\sin\theta_d)$ &
$g_q(V_{td}\lambda_{ds}+V_{tb}\lambda_{sb}+V_{ts}\lambda_{ss})/\sqrt{2}$\\  \hline
\end{tabular}
\end{center}
\end{table}

 \begin{table}[h]
 \begin{center}
 \caption{\label{tab2} Flavor conserving couplings to $Z^\prime$ boson in different parameterizations of the TVB model.}
\begin{tabular}{|c|c|c|c|}
\hline
 Coupling & Parameterization in \cite{Kumar:2018kmr} &  Parameterization in \cite{Calibbi:2015kma,Bhattacharya:2016mcc}& Parameterization in \cite{Greljo:2015mma,Buttazzo:2017ixm}\\ \hline
 $g_{uu}^q$ & $g_{bb}V_{ub}^2+2g_{sb}V_{us}V_{ub}+g_{ss}V_{us}^2$ & $g_2^q(V_{ub}^2\cos^2\theta_d-2V_{us}V_{ub}\cos\theta_d\sin\theta_d+V_{us}^2\sin^2\theta_d)$ & $g_q\lambda_{uu}/\sqrt{2}$\\
 $g_{cc}^q$ & $g_{bb}V_{cb}^2+2g_{sb}V_{cs}V_{cb}+g_{ss}V_{cs}^2$ &
$g_2^q(V_{cb}^2\cos^2\theta_d-2V_{cs}V_{cb}\cos\theta_d\sin\theta_d+V_{cs}^2\sin^2\theta_d)$&$g_q\lambda_{cc}/\sqrt{2}$\\
 $g_{tt}^q$ & $g_{bb}V_{tb}^2+2g_{sb}V_{ts}V_{tb}+g_{ss}V_{ts}^2$ &
$g_2^q(V_{tb}^2\cos^2\theta_d-2V_{ts}V_{tb}\cos\theta_d\sin\theta_d+V_{ts}^2\sin^2\theta_d)$&$g_q\lambda_{tt}/\sqrt{2}$\\
\hline
\end{tabular}
\end{center}

 \end{table}
 
 \begin{table}[h]
     \begin{center}
      \caption{\label{tab3}Flavor changing couplings to $Z^\prime$ boson in different parameterizations of the TVB model.}
    \begin{tabular}{|c|c|c| c|}
    \hline
 Coupling & Parameterization in \cite{Kumar:2018kmr} &  Parameterization in \cite{Calibbi:2015kma,Bhattacharya:2016mcc}& Parameterization in \cite{Greljo:2015mma,Buttazzo:2017ixm}\\ \hline
 $g_{uc}^q$ & $ g_{bb}V_{cb}V_{ub}+g_{sb}V_{cs}V_{ub}$ & $ g_2^q V_{cb}V_{ub}\cos^2\theta_d- g_2^q V_{cs}V_{ub}\cos\theta_d\sin\theta_d$ & $g_q\lambda_{uc}/\sqrt{2}$\\
  &$+g_{sb}V_{cb}V_{us}+ g_{ss}V_{cs}V_{us}$ & $- g_2^q V_{cb}V_{us}\cos\theta_d\sin\theta_d+g_2^q V_{cs}V_{us}\sin^2\theta_d$& \\
  $g_{ut}^q$ & $ g_{bb}V_{tb}V_{ub}+g_{sb}V_{ts}V_{ub}$ & $ g_2^q V_{tb}V_{ub}\cos^2\theta_d- g_2^q V_{ts}V_{ub}\cos\theta_d\sin\theta_d$& $g_q\lambda_{ut}/\sqrt{2}$\\
  &$+g_{sb}V_{tb}V_{us}+ g_{ss}V_{ts}V_{us}$ & $- g_2^q V_{tb}V_{us}\cos\theta_d\sin\theta_d+g_2^q V_{ts}V_{us}\sin^2\theta_d$&\\
  $g_{ct}^q$ & $ g_{bb}V_{cb}V_{tb}+g_{sb}V_{cs}V_{tb}$ & $ g_2^q V_{cb}V_{tb}\cos^2\theta_d- g_2^q V_{cs}V_{tb}\cos\theta_d\sin\theta_d$&$g_q\lambda_{ct}/\sqrt{2}$\\
  &$+g_{sb}V_{cb}V_{ts}+ g_{ss}V_{cs}V_{ts}$ & $- g_2^q V_{cb}V_{ts}\cos\theta_d\sin\theta_d+g_2^q V_{cs}V_{ts}\sin^2\theta_d$ &\\
\hline 
    \end{tabular}
\end{center}
 \end{table}

 \begin{table}[h!]
     \begin{center}
      \caption{\label{tab4}Couplings of leptons to $Z^\prime$ boson in different parameterizations of the TVB model.}
    \begin{tabular}{|c|c|c| c|}
    \hline
 Coupling & Parameterization in \cite{Kumar:2018kmr}  &  Parameterization in \cite{Calibbi:2015kma,Bhattacharya:2016mcc}& Parameterization in \cite{Greljo:2015mma,Buttazzo:2017ixm}\\ \hline
$g_{\mu\mu}$ & $g_{\mu\mu}$ & $g_2^\ell\sin^2\theta_L$ &
$g_q(\lambda_{\mu\mu})/\sqrt{2}$\\
$g_{\mu\tau}$ & $g_{\mu\tau}$ & $-g_2^\ell\sin\theta_L\cos\theta_L$ &
$2g_q(\lambda_{\mu\tau})/\sqrt{2}$\\
$g_{\tau\tau}$ & $g_{\tau\tau}$ & $ g_2^\ell\cos^2\theta_L$ &
$g_q/\sqrt{2}$\\
\hline 
    \end{tabular}
\end{center}
 \end{table}

 We make emphasis that the results presented in tables \ref{tab1}, \ref{tab2},~\ref{tab3}, and~\ref{tab4} allow us to understand the differences and similarities for the parameterizations presented above in the context of the TVB model; additionally it gives us a complete interpretation of the variables present on each one and the possibilities to find adjustments to explain flavor anomalies.

\section{Relevant Observables} \label{Obs}

In this section, we discuss the constraints from the most relevant flavor observables on the TVB model couplings that simultaneously accommodate the $B$ meson anomalies. We will include the recent experimental progress from Belle and LHCb on different LFV decays (such as $\Upsilon(1S) \to \mu^\pm \tau^\mp$, $B \to K^{\ast} \mu^\pm \tau^\mp$, and $\tau \to \mu\phi$).

\subsection{$b \to c \ell^- \bar{\nu}_\ell$ ($\ell = \mu, \tau$) data}

The $W^\prime$ boson leads to additional tree-level contribution to $b \to c \ell^- \bar{\nu}_\ell$ transitions involving leptons from second- and third-generation $(\ell=\mu,\tau)$. The total low-energy effective Lagrangian has the following form~\cite{Gomez:2019xfw}
\bea\label{Leff_Total}
- \mathcal{L}_{\rm eff}(b \to c \ell \bar{\nu}_{\ell})_{\rm SM+W^\prime} &=& \frac{4 G_F}{\sqrt{2}} V_{cb}	 \Big[(1+ C_{V}^{bc\ell\nu_\ell})(\bar{c} \gamma_\mu P_L b) (\bar{\ell} \gamma^\mu P_L \nu_{\ell})  \Big], 
\eea
\noindent where $G_F$ is the Fermi coupling constant, $V_{cb}$, is the charm-bottom Cabbibo-Kobayashi-Maskawa (CKM) matrix element, and $C_{V}^{bc\ell\nu_\ell}$ is the Wilson coefficient (WC) associated with the NP vector (left-left) operator. This WC is defined as  
\begin{eqnarray}\label{C_V}
C_{V}^{bc\ell\nu_\ell} &=& \frac{\sqrt{2}}{4 G_F V_{cb}} \frac{2(V_{cs} g^q_{sb} + V_{cb} g^q_{bb}) g^\ell_{\ell\ell}}{M_{V}^2}  \ \ \ (\ell=\mu,\tau),
\end{eqnarray}

\noindent with $M_{V}$ the heavy boson mass. The NP effects on the LFU ratios $R(X)$ ($X=D,D^\ast,J/\psi$), the $D^\ast$ and $\tau$ longitudinal polarizations related with the channel $\bar{B} \to D^\ast \tau\bar{\nu}_\tau$, the ratio of inclusive decays $R(X_c)$, and the tauonic decay $B_c^- \to \tau^- \bar{\nu}_{\tau}$ can be easily parametrized as~\cite{Gomez:2019xfw}
\begin{eqnarray}
R(X) &=& R(X)_{\rm SM} \big| 1 + C_{V}^{bc\tau\nu_\tau} \big|^2,  \label{RD} \\
F_L(D^*) &=&  F_L(D^*)_{\rm SM} \ r_{D^\ast}^{-1}   \big| 1 + C_{V}^{bc\tau\nu_\tau}\big|^2 , \label{FLD} \\
P_\tau(D^*) &=&  P_\tau(D^*)_{\rm SM} \ r_{D^\ast}^{-1} \big\vert 1 + C_{V}^{bc\tau\nu_\tau} \big\vert^2 \ , \label{PTAU} \\
R(X_c) &=& R(X_c)_{\rm SM} \Big( 1+ 2.294 \ {\rm Re}(C_{V}^{bc\tau\nu_\tau}) + 1.147 \big\vert C_{V}^{bc\tau\nu_\tau} \big \vert^2 \Big ), \label{R_Xc} \\
{\rm BR}(B_c^- \to \tau^- \bar{\nu}_{\tau}) &=& {\rm BR}(B_c^- \to \tau^- \bar{\nu}_{\tau})_{\text{SM}}   \big\vert 1+C_{V}^{bc\tau\nu_\tau} \big\vert^2  \label{BRBc_modified},
\end{eqnarray}

\noindent respectively, where $r_{D^\ast} = R(D^*) / R(D^*)_{\rm SM}$. For ${\rm BR}(B_c^- \to \tau^- \bar{\nu}_{\tau})$, we will use the bound $< 10\%$~\cite{Akeroyd:2017mhr}. Concerning to the ratio $R(\Lambda_c)$ very recently measured by LHCb~\cite{LHCb:2022piu}, the SM contribution is also rescaled by the overall factor $\big| 1 + C_{V}^{bc\tau\nu_\tau} \big|^2$, namely~\cite{Datta:2017aue}
\begin{equation}
R(\Lambda_c) = R(\Lambda_c)_{\rm SM} \ \big| 1 + C_{V}^{bc\tau\nu_\tau} \big|^2.
\end{equation}

\noindent A long term  integrated luminosity of $50 \ {\rm ab}^{-1}$ is expected to be accumulated by the Belle II experiment~\cite{Belle-II:2018jsg}, allowing improvements at the level of $\sim 3\%$ and  $\sim 2\%$ for the statistical and systematic uncertainties of $R(D)$ and $R(D^{\ast})$, respectively~\cite{Belle-II:2018jsg}. It is also envisioned accuracy improvements on angular analysis in $\bar{B} \to D^\ast \tau\bar{\nu}_\tau$ decay ($\tau$ polarization observable $P_\tau(D^*)$), as well as on $q^2$-distribution~\cite{Belle-II:2018jsg}. On the other hand, the LHCb will be able to improve measurements of $R(D^{\ast})$ and $R(J/\psi)$ in the future runs of data taking~\cite{Albrecht:2021tul,Bifani:2018zmi}.

In regard to the transition $b\to c \mu \bar{\nu}_\mu$, the $\mu/e$ LFU ratios $R_{D^{(\ast)}}^{\mu/e} \equiv {\rm BR}(B\to D^{(\ast)}\mu \bar{\nu}_\mu)/{\rm BR}(B\to D^{(\ast)} e \bar{\nu}_e)$ have to be taken into account. The experimental values obtained by Belle~\cite{Glattauer:2015teq,Belle:2017rcc} are in great accordance with the SM estimations~\cite{Becirevic:2020rzi,Bobeth:2021lya} (see Table~\ref{Table:1}). The $W^\prime$ boson coupling to lepton pair $\mu \bar{\nu}_\mu$ modifies this ratio as 
\begin{equation}
R_{D^{(\ast)}}^{\mu/e} = [R_{D^{(\ast)}}^{\mu/e}]_{\rm SM}\big|1+C_{V}^{bc\mu\nu_\mu} \big|^2,
\end{equation}

\noindent where $C_{V}^{bc\mu\nu_\mu}$ is given by Eq.~\eqref{C_V}. From this LFU ratio we get the bound
\begin{equation}
\frac{\vert(V_{cs} g^q_{sb} + V_{cb} g^q_{bb}) g^\ell_{\mu\mu}\vert}{M_{V}^2} \leqslant 0.013 \ {\rm TeV}^{-2},
\end{equation}
which is relevant for the couplings aiming to explain the $b \to s\mu^+\mu^-$ anomaly (see Sec.~\ref{bsmupmum}).

\subsection{$b \to u \ell^- \bar{\nu}_\ell$ ($\ell=\mu,\tau$) data}

The TVB model can also induce NP contributions in the leptonic decay $B\to \ell\bar{\nu}_\ell$ induced via the charged-current transition $b \to u \ell^- \bar{\nu}_\ell$ ($\ell=\mu,\tau$). The ratio 
\begin{equation}
R_B^{\tau/\mu} \equiv \dfrac{{\rm BR}(B^- \to \tau^- \bar{\nu}_{\tau})}{{\rm BR}(B^- \to \mu^- \bar{\nu}_{\mu})},
\end{equation}
provides a clean LFU test~\cite{Becirevic:2020rzi}. Through this ratio the uncertainties on the decay constant $f_B$ and CKM element $V_{ub}$ cancel out (circumventing the tension between the exclusive and inclusive values of $V_{ub}$~\cite{UTfit:2022hsi}). The NP effects on this ratio can be expressed as
\begin{equation}
R_B^{\tau/\mu} = [R_B^{\tau/\mu}]_{\rm SM} \Bigg\vert\dfrac{ 1+C_{V}^{bu\tau\nu_\tau}}{1+C_{V}^{bu\mu\nu_\mu} }\Bigg\vert^2,
\end{equation}
\noindent where
\begin{equation}
C_{V}^{bu\ell\nu_\ell} =  \frac{\sqrt{2}}{4 G_F  M_{U_1}^2} \Big[\vert x_L^{b\ell} \vert^2 + \frac{V_{us}}{V_{ub}} x_L^{s\tau} (x_L^{b\ell})^\ast \Big], \ \ (\ell = \mu,\tau)
\end{equation}
and 
\begin{equation}
[R_B^{\tau/\mu}]_{\rm SM} = \Big( \dfrac{m_{\tau}}{m_{\mu}}\Big)^2 \Big( \dfrac{m_{B}^2 - m_{\tau}^2}{m_{B}^2 - m_{\mu}^2}\Big)^2 = 222.5 \pm 3.0.
\end{equation}
The experimental value is $[R_B^{\tau/\mu}]_{\rm Exp} = 205.7 \pm 96.6$, which was obtained from the values reported by the Particle Data Group (PDG) on ${\rm BR}(B^- \to \tau^- \bar{\nu}_{\tau})$~\cite{PDG2020} and the Belle experiment on ${\rm BR}(B^- \to \mu^- \bar{\nu}_{\mu})$~\cite{Belle:2019iji}.

\subsection{$b \to s \mu^+\mu^-$ data} 
\label{bsmupmum}

The NP effective Lagrangian responsible for the semileptonic transition $b \to s \mu^+\mu^-$ can be expressed as
\begin{equation}
\mathcal{L}(b \to s \mu^+\mu^-)_{\rm NP} = \frac{4G_F}{\sqrt{2}} V_{tb} V_{ts}^\ast (C^{bs\mu\mu}_9 \mathcal{O}^{bs\mu\mu}_9 + C^{bs\mu\mu}_{10} \mathcal{O}^{bs\mu\mu}_{10}) + \ {\rm h.c.},
\end{equation}

\noindent where the NP is encoded in the WCs $C^{bs\mu\mu}_{9}$ and $C^{bs\mu\mu}_{10} $ of the four-fermion operators
\begin{eqnarray}
\mathcal{O}^{bs\mu\mu}_9 &=& \frac{\alpha_{\rm em}}{4\pi} (\bar{s} \gamma_\mu P_{L} b) (\bar{\mu}\gamma^\mu \mu), \\
\mathcal{O}^{bs\mu\mu}_{10} &=& \frac{\alpha_{\rm em}}{4\pi} (\bar{s} \gamma_\mu P_{L} b) (\bar{\mu}\gamma^\mu \gamma_5 \mu),
\end{eqnarray}

\noindent respectively, with $\alpha_{\rm em}$ being the fine-constant structure. A global fit analysis including most current $b \to s \mu^+\mu^-$ data, such as $R_{K^{(\ast)}}$ by LHCb~\cite{LHCb:2022qnv,LHCb:2022zom} and ${\rm BR}(B_s \to \mu^+\mu^-)$ by CMS~\cite{CMS:2022mgd}, has been recently performed in Ref.~\cite{Greljo:2022jac,Alguero:2023jeh}. Among the different NP scenarios, the $C^{bs\mu\mu}_{9} = - C^{bs\mu\mu}_{10}$ solution is preferred by the data~\cite{Greljo:2022jac,Alguero:2023jeh}.\footnote{Let us notice that the single WC $C^{bs\mu\mu}_{9}$ also provides a good fit of the $b \to s \mu^+\mu^-$ data~\cite{Greljo:2022jac,Alguero:2023jeh}. Some explicit model examples are shown in~\cite{Greljo:2022jac}.} The best fit $1\sigma$ solution is~\cite{Greljo:2022jac}
\begin{equation} \label{C9}
C^{bs\mu\mu}_{9} = - C^{bs\mu\mu}_{10} \in [-0.23,-0.11].
\end{equation}

\noindent In the context of the TVB model, the $Z^\prime$ boson induces a tree-level contribution to $b \to s \mu^+\mu^-$ transition via the WCs
\begin{equation}
C^{bs\mu\mu}_{9} = - C^{bs\mu\mu}_{10} = - \frac{\pi}{\sqrt{2} G_F \alpha_{\rm em} V_{tb} V_{ts}^\ast} \frac{g_{sb}^q g_{\mu\mu}^{\ell}}{M_{V}^2}.
\end{equation}

\noindent Using the result of the global fit, Eq.~\eqref{C9}, this corresponds to 
\begin{equation}
 -\frac{g_{sb}^q g_{\mu\mu}^{\ell}}{M_{V}^2} \in [1.7, 3.5] \times 10^{-4} \ {\rm TeV}^{-2}.
\end{equation}

\subsection{Bottomonium processes: $R_{\Upsilon(nS)}$ and $\Upsilon(nS) \to \mu^\pm \tau^\mp$}

Test of LFU has also been studied in the leptonic ratio $R_{\Upsilon(nS)}$ (with $n=1,2,3$) in connection with the reported hints of LFU violation in the charged-current transition $b \to c \tau \bar{\nu}_{\tau}$~\cite{Aloni:2017eny,Garcia-Duque:2021qmg}.\footnote{Recently, in Ref.~\cite{Descotes-Genon:2021uez} has been proposed a new method to test LFU through inclusive dileptonic $\Upsilon(4S)$ decays.} It is known that NP scenarios aiming to provide an explanation to the anomalous $b \to c \tau \bar{\nu}_{\tau}$ data, also induce effects in the neutral-current transition $b\bar{b} \to  \tau^{+}\tau^{-}$~\cite{Aloni:2017eny,Garcia-Duque:2021qmg}. Experimentally, the BABAR and CLEO Collaborations have reported the values~\cite{delAmoSanchez:2010bt,Besson:2006gj,Lees:2020kom}
\begin{eqnarray}
R_{\Upsilon(1S)} &=&
\begin{cases}
\text{BABAR-10:} \ 1.005 \pm 0.013 \pm 0.022 \text{~\cite{delAmoSanchez:2010bt}}, \\
\text{SM:} \ 0.9924 \text{~\cite{Aloni:2017eny}},
\end{cases} \\
R_{\Upsilon(2S)} &=&
\begin{cases}
\text{CLEO-07:} \ 1.04 \pm 0.04 \pm 0.05 \text{~\cite{Besson:2006gj}}, \\
\text{SM:} \ 0.9940 \text{~\cite{Aloni:2017eny}},
\end{cases} \\
R_{\Upsilon(3S)} &=& 
\begin{cases}
\text{CLEO-07:} \ 1.05 \pm 0.08 \pm 0.05 \text{~\cite{Besson:2006gj}}, \\
\text{BABAR-20:} \ 0.966 \pm 0.008 \pm 0.014 \text{~\cite{Lees:2020kom}}, \\
\text{SM:} \ 0.9948 \text{~\cite{Aloni:2017eny}},
\end{cases} 
\end{eqnarray} 

\noindent where the theoretical uncertainty is typically of the order $\pm \mathcal{O}(10^{-5})$~\cite{Aloni:2017eny}. These measurements are in good accordance with the SM estimations, except for the 2020 measurement on $R_{\Upsilon(3S)}$ that shows an agreement at the $1.8\sigma$ level~\cite{Lees:2020kom}. By averaging the CLEO-07~\cite{Besson:2006gj} and BABAR-20~\cite{Lees:2020kom} measurements we obtain $R_{\Upsilon(3S)}^{\rm Ave} = 0.968 \pm 0.016$, which deviates at the $1.7\sigma$ level with respect to the SM prediction~\cite{Garcia-Duque:2021qmg}.

The NP effects of the TVB model on the leptonic ratio can be expressed as~\cite{Aloni:2017eny,Garcia-Duque:2021qmg}
\begin{equation}
R_{\Upsilon(nS)}= \frac{(1-4x_\tau^2)^{1/2}}{\vert A_V^{\rm SM} \vert^2} \Big[ \vert A_V^{b\tau} \vert^2 (1+2x_\tau^2) + \vert B_V^{b\tau} \vert^2 (1- 4x_\tau^2) \Big],
\end{equation}

\noindent with $x_\tau = m_\tau/m_{\Upsilon(nS)}$, $\vert A_V^{\rm SM} \vert = - 4\pi\alpha Q_b$, and 
\begin{eqnarray}
A_V^{b\tau} &=& - 4\pi\alpha Q_b + \frac{m_{\Upsilon(nS)}^2}{4} \frac{g^q_{bb} g^\ell_{\tau\tau}}{4M_{V}^2}, \\
B_V^{b\tau} &=&- \frac{m_{\Upsilon(nS)}^2}{2} \frac{g^q_{bb} g^\ell_{\tau\tau}}{4M_{V}^2}.
\end{eqnarray}

The neutral gauge boson also generates the LFV processes $\Upsilon \to \mu^\pm \tau^\mp$ $(\Upsilon \equiv \Upsilon(nS))$. The branching fraction is given by~\cite{Bhattacharya:2016mcc,Kumar:2018kmr}
\begin{equation}
{\rm BR}(\Upsilon \to \mu^\pm \tau^\mp) =  \frac{f_\Upsilon^2 m_\Upsilon^3}{48\pi \Gamma_\Upsilon} \Big(2+ \frac{m_\tau^2}{m_\Upsilon^2} \Big) \Big(1- \frac{m_\tau^2}{m_\Upsilon^2} \Big)^2 \Big\vert \dfrac{g^q_{bb} (g^\ell_{\mu\tau})^\ast}{M_{V}^2} \Big\vert^2,
\end{equation}

\noindent where $f_{\Upsilon}$ and $m_{\Upsilon}$ are the Upsilon decay constant and mass, respectively. The decay constant values can be extracted from the experimental branching ratio measurements of the processes $\Upsilon \to e^-e^+$. Using current data from PDG~\cite{PDG2020}, one obtains $f_{\Upsilon(1S)} = (659 \pm 17) \ {\rm MeV}$,  $f_{\Upsilon(2S)} = (468 \pm 27) \ {\rm MeV}$, and $f_{\Upsilon(3S)} = (405 \pm 26) \ {\rm MeV}$. Experimentally, the reported ULs are ${\rm BR}(\Upsilon(1S) \to \mu^\pm \tau^\mp)  < 2.7 \times 10^{-6}$ from Belle~\cite{Belle:2022cce}, and ${\rm BR}(\Upsilon(2S) \to \mu^\pm \tau^\mp)  < 3.3 \times 10^{-6}$, ${\rm BR}(\Upsilon(3S) \to \mu^\pm \tau^\mp)  < 3.1 \times 10^{-6}$ from PDG~\cite{PDG2020}. From these ULs we get
\begin{subequations}
\begin{eqnarray}
  \Upsilon(1S) \to \mu^\pm \tau^\mp &:&  \  \frac{\vert g^q_{bb}(g^\ell_{\mu\tau})^\ast\vert}{M_{V}^2}  < 5.7  \ {\rm TeV}^{-2}, \\
   \Upsilon(2S) \to \mu^\pm \tau^\mp &:& \ \frac{\vert g^q_{bb}(g^\ell_{\tau\mu})^\ast\vert}{M_{V}^2}  < 6.2  \ {\rm TeV}^{-2},  \\
   \Upsilon(3S) \to \mu^\pm \tau^\mp &:& \ \frac{\vert g^q_{bb}(g^\ell_{\mu\tau})^\ast\vert}{M_{V}^2}  < 5.2 \ {\rm TeV}^{-2}.
\end{eqnarray}
\end{subequations}


\subsection{$\Delta F = 2$ processes: $B_s-\bar{B}_s$ and $D^0 -\bar{D}^0$mixing}
\label{mixing}

The interactions of a $Z^\prime$ boson to quarks $s\bar{b}$ relevant for $b \to s \mu^+\mu^-$ processes also generate a contribution to $B_s-\bar{B}_s$ mixing~\cite{DiLuzio:2019jyq,DiLuzio:2017fdq}. The NP effects to the $B_s-\bar{B}_s$ mixing can be described by the effective Lagrangian
\begin{equation}
\mathcal{L}_{\rm \Delta B = 2}^{Z^\prime} = - \frac{4G_F}{\sqrt{2}} \vert V_{tb} V_{ts}^\ast \vert^2 C_{sb}^{LL} (\bar{s} \gamma_\mu  P_L b) (\bar{s} \gamma^\mu  P_L b) + \ {\rm h.c.},
\end{equation}

\noindent where 
\begin{equation}
C_{sb}^{LL} = \frac{1}{4\sqrt{2} G_F \vert V_{tb} V_{ts}^\ast \vert^2} \frac{\vert g_{sb}^q \vert^2}{M_{Z^\prime}^2}.
\end{equation}

\noindent Thus, the NP contributions to the mass difference $\Delta M_s$ of the neutral $B_s$ meson can be expressed as~\cite{DiLuzio:2019jyq}
\begin{equation}
\dfrac{\Delta M_s^{\rm SM+NP}}{\Delta M_s^{\rm SM}} = \Big(1+ \frac{\eta^{6/23}}{R^{\rm loop}_{\rm SM}} C_{sb}^{LL} \Big),
\end{equation}

\noindent where $\eta = \alpha_s(M_{Z^\prime})/\alpha_s(m_b)$ accounts for running from the $M_{Z^\prime}$ scale down to the $b$-quark mass scale and the SM loop function is $R^{\rm loop}_{\rm SM} = (1.310 \pm 0.010)\times 10^{-3}$~\cite{DiLuzio:2019jyq}. At present, $\Delta M_s$ has been experimentally measured with great precision $\Delta M_s^{\rm Exp} = (17.757 \pm 0.021) \ {\rm ps}^{-1}$~\cite{DiLuzio:2019jyq,HFLAV:2022pwe}. On the theoretical side, the average is $\Delta M_s^{\rm SM} = (18.4^{+0.7}_{-1.2}) \ {\rm ps}^{-1}$ implying that $\Delta M_s^{\rm SM} / \Delta M_s^{\rm Exp} = 1.04^{+0.04}_{-0.07}$~\cite{DiLuzio:2019jyq}. This value yields to 
\begin{equation}
0.89 \leq \Bigg\vert 1+ \frac{\eta^{6/23}}{R_{\rm loop}^{\rm SM}} C_{sb}^{LL} \Bigg\vert \leq 1.11,
\end{equation}
where in the TVB model translates into the important $2\sigma$ bound
\begin{equation}
\frac{\vert g_{sb}^q \vert}{M_{V}} \geq 3.9 \times 10^{-3} \ {\rm TeV^{-1}}.
\end{equation} 

In addition, the $Z^\prime$ boson can also admit $c\to u$ transitions, consequently generating tree-level effects on $D^0 - \bar{D}^0$ mixing~\cite{Kumar:2018kmr,Alok:2021pdh}. The effective Lagrangian describing the $Z^\prime$ contribution to $D^0 - \bar{D}^0$ mixing can be expressed as~\cite{Kumar:2018kmr,Alok:2021pdh}
\begin{equation}
\mathcal{L}_{\rm \Delta C = 2}^{Z^\prime} = - \frac{\vert g_{uc} \vert^2}{2M_{Z^\prime}^2} (\bar{c} \gamma_\mu P_L u) (\bar{c} \gamma^\mu P_L u) + \ {\rm h.c.},
\end{equation}

\noindent where $g_{uc} = g^q_{bb} V_{cb}V^\ast_{ub} + g^q_{sb} (V_{cs}V^\ast_{ub} + V_{cb}V^\ast_{us}) + g^q_{ss}V_{cs}V^\ast_{us}$~\cite{Kumar:2018kmr} (see also Table~\ref{tab3}). Such a NP contributions are constrained by the results of the mass difference $\Delta M_D$ of neutral $D$ mesons. The theoretical determination of this mass difference is limited by our understanding of the short and long-distance contributions~\cite{Kumar:2018kmr,Alok:2021pdh}. Here we follow the recent analysis of Ref.~\cite{Kumar:2018kmr} focused on short-distance SM contribution that sets the conservative (strong) bound
\begin{equation}
\frac{\vert g_{ss}^q \vert}{M_{V}} \leq 3 \times 10^{-3} \ {\rm TeV^{-1}}.
\end{equation}

\noindent The couplings $g^q_{bb}$ and $g^q_{sb}$ are less constrained by $\Delta M_D$~\cite{Kumar:2018kmr}, therefore, we will skip them in our study.

\subsection{Neutrino Trident Production}
The $Z^\prime$ couplings to leptons from second-generation ($g_{\mu\mu}=g_{\nu_\mu \nu_\mu}$) also generate a contribution to the cross-section of neutrino trident production (NTP), $\nu_\mu N \to \nu_\mu N \mu^+\mu^-$~\cite{Altmannshofer:2014pba}. The cross-section is given by~\cite{Altmannshofer:2014pba}
\begin{equation}
\frac{\sigma_{\rm SM+NP}}{\sigma_{\rm SM}}= \frac{1}{1+(1+4s_W^2)^2} \Big[ \Big(1+\frac{v^2g_{\mu\mu}^2}{M_V^2}\Big)^2 + \Big(1+4s_W^2+\frac{v^2g_{\mu\mu}^2}{M_V^2}\Big)^2\Big], 
\end{equation}

\noindent where $v= (\sqrt{2}G_F)^{-1/2}$ and $s_W \equiv \sin\theta_W$ (with $\theta_W$ the Weinberg angle). The existing CCFR trident measurement $\sigma_{\rm CCFR}/\sigma_{\rm SM} = 0.82\pm 0.28$ provides the upper bound
\begin{equation}
\frac{\vert g_{\mu\mu}^\ell \vert}{M_{Z^\prime}} \leq 1.13 \ {\rm TeV^{-1}}.
\end{equation}

\subsection{LFV $B$ decays: $B\to K^{(\ast)} \mu^\pm \tau^\mp$ and $B_s \to \mu^\pm \tau^\mp$}

The $Z^\prime$ boson mediates LFV transitions $b \to s\mu^\pm\tau^\mp$ ($B \to K^{(\ast)} \mu^\pm \tau^\mp$ and $B_s^0 \to \mu^\pm \tau^\mp$) at tree level via the WCs~\cite{Calibbi:2015kma}
\begin{equation}
C^{bs\mu\tau}_{9} = - C^{bs\mu\tau}_{10} = - \frac{\pi}{\sqrt{2} G_F \alpha_{\rm em} V_{tb} V_{ts}^\ast} \frac{g^q_{sb}(g^\ell_{\mu\tau})^\ast}{M_{V}^2}.
\end{equation}

\noindent The current experimental limits ($90\%$ C.L.) on the branching ratios of $B^+\to K^+ \mu^\pm \tau^\mp$ are~\cite{PDG2020}
\begin{eqnarray}
{\rm BR}(B^+ \to K^+\mu^+\tau^-)_{\rm exp} &<& 4.5 \times 10^{-5}, \\
{\rm BR}(B^+ \to K^+\mu^-\tau^+)_{\rm exp} &<& 2.8 \times 10^{-5}.
\end{eqnarray}

\noindent Let us notice that LHCb Collaboration obtained a limit of ${\rm BR}(B^+ \to K^+\mu^-\tau^+)_{\rm LHCb} < 3.9 \times 10^{-5}$~\cite{Aaij:2020mqb} that is comparable with the one quoted above from PDG. On the other hand, the LHCb has recently presented the first search of $B^0 \to K^{\ast 0}\mu^\pm\tau^\mp$~\cite{LHCb:2022wrs}. The obtained UL on this LFV decay is~\cite{LHCb:2022wrs}
\begin{equation}
{\rm BR}(B^0 \to K^{\ast 0}\mu^\pm\tau^\mp)_{\rm exp} < 1.0 \times 10^{-5}.
\end{equation}

\noindent From the theoretical side, the branching ratio of $B^+ \to K^+\mu^+\tau^-$~\cite{Parrott:2022zte} and $B^0 \to K^{\ast 0}\mu^+\tau^-$~\cite{Calibbi:2015kma} can be written as
\begin{eqnarray}
{\rm BR}(B^+ \to K^+\mu^+\tau^-) &=& \big(a_{K} \vert C^{bs\mu\tau}_{9} \vert^2+ b_{K} \vert C^{bs\mu\tau}_{10} \vert^2 \big) \times 10^{-9} , \\
{\rm BR}(B^0 \to K^{\ast 0}\mu^+\tau^-) &=& \Big( (a_{K^\ast}+c_{K^\ast}) \vert C^{bs\mu\tau}_{9} \vert^2 + (b_{K^\ast}+d_{K^\ast}) \vert C^{bs\mu\tau}_{10} \vert^2 \Big) \times 10^{-9} ,
\end{eqnarray}

\noindent respectively, where $(a_{K},b_{K}) = (12.72 \pm 0.81, 13.21 \pm 0.81)$~\cite{Parrott:2022zte}, and $(a_{K^\ast},b_{K^\ast},c_{K^\ast},d_{K^\ast}) =(3.0\pm 0.8, 2.7\pm 0.7,16.4\pm 2.1, 15.4\pm 1.9)$~\cite{Calibbi:2015kma} are the numerical coefficients that have been calculated using the $B\to K^{(\ast)}$ transitions form factors obtained from lattice QCD~\cite{Parrott:2022zte,Calibbi:2015kma}. The decay channel with final state $\mu^-\tau^+$ can be easily obtained by replacing $\mu \leftrightarrows \tau$. The current ULs can be translated into the bounds
\begin{subequations}
\begin{eqnarray}
  B^+ \to K^+\mu^+\tau^- &:&  \  \frac{\vert g^q_{sb}(g^\ell_{\mu\tau})^\ast\vert}{M_{V}^2}  < 6.2\times 10^{-2}  \ {\rm TeV}^{-2}, \\
   B^+ \to K^+\mu^-\tau^+ &:& \ \frac{\vert g^q_{sb}(g^\ell_{\tau\mu})^\ast\vert}{M_{V}^2}  < 4.9\times 10^{-2}  \ {\rm TeV}^{-2},  \\
   B^0 \to K^{\ast 0}\mu^+\tau^- &:& \ \frac{\vert g^q_{sb}(g^\ell_{\mu\tau})^\ast\vert}{M_{V}^2}  < 2.5\times 10^{-2}  \ {\rm TeV}^{-2}.
\end{eqnarray}
\end{subequations}


As for the LFV leptonic decay $B_s \to \mu^\pm \tau^\mp$, the branching ratio is~\cite{Calibbi:2015kma}
\begin{eqnarray}
{\rm BR}(B_s^0 \to \mu^\pm \tau^\mp) &=& \tau_{B_s}\frac{f_{B_s}^2 m_{B_s} m^2_\tau}{32\pi^3} \alpha^2 G_F^2 \vert V_{tb} V_{ts}^\ast \vert^2 \Big(1-  \frac{m_\tau^2}{m_{B_s}^2} \Big)^2 \big(\vert C^{bs\mu\tau}_{9} \vert^2+  \vert C^{bs\mu\tau}_{10} \vert^2 \big), 
\end{eqnarray}

\noindent where $f_{B_s} = (230.3 \pm 1.3)$ MeV is the $B_s$ decay constant~\cite{HFLAV:2022pwe} and  we have used the limit $m_\tau \gg m_\mu$. Recently, the LHCb experiment has reported the first upper limit of ${\rm BR}(B_s \to \mu^\pm \tau^\mp) <4.2 \times 10^{-5}$ at $95\%$ CL~\cite{Aaij:2019okb}. Thus, one gets the following limit
\begin{equation}
  \frac{\vert g^q_{sb}(g^\ell_{\mu\tau})^\ast\vert}{M_{V}^2}  <  5.1 \times 10^{-2}  \ {\rm TeV}^{-2}.
\end{equation}

\subsection{Rare $B$ decays: $B \to K^{(\ast)} \nu\bar{\nu}$, $B\to K \tau^+\tau^-$ and $B_s\to \tau^+\tau^-$}

Recently, the interplay between the di-neutrino channel $B \to K^{(\ast)} \nu\bar{\nu}$ and the $B$ meson anomalies has been studied by several works~\cite{Alok:2021pdh,Bause:2020auq,Bause:2021cna,Browder:2021hbl,He:2021yoz}. In the NP scenario under study, the $Z^\prime$ boson can give rise to $B \to K^{(\ast)} \nu\bar{\nu}$ at tree level. The effective Hamiltonian for the $b \to s \nu\bar{\nu}$ transition is given by~\cite{Buras:2014fpa}
\begin{equation}
\mathcal{H}_{\rm eff}(b \to s \nu \bar{\nu}) = - \frac{\alpha_{\rm em}G_F}{\sqrt{2}\pi}V_{tb}V_{ts}^\ast  C_L^{ij} (\bar{s}\gamma^\mu P_L b)(\bar{\nu}_i \gamma_\mu(1-\gamma_5) \nu_j),
\end{equation}

\noindent where $C_L^{ij} = C_L^{\rm SM} + \Delta C_L^{ij}$ is the aggregate of the SM contribution $C_L^{\rm SM} \approx -6.4$ and the NP effects $\Delta C_L^{ij}$, that in the TVB framework read as
\begin{equation} \label{C_nu}
\Delta C_L^{ij}  = \frac{\pi}{\sqrt{2} G_F \alpha_{\rm em} V_{tb} V_{ts}^\ast} \frac{g^q_{sb}g^\ell_{ij}}{M_{V}^2},
\end{equation}

\noindent with $i,j = \mu, \tau$. By defining the ratio~\cite{Buras:2014fpa}
\begin{equation}
R^{\nu\bar{\nu}}_{K^{(\ast)}} \equiv \frac{{\rm BR}(B \to K^{(\ast)} \nu\bar{\nu})}{{\rm BR}(B \to K^{(\ast)} \nu\bar{\nu})_{\rm SM}},
\end{equation}
the NP contributions can be constrained. In the TVB model this ratio is modified as
\begin{eqnarray}
R^{\nu\bar{\nu}}_{K^{(\ast)}}  &=&  \frac{\sum_{ij}\vert \delta_{ij} C_L^{\rm SM} + \Delta C_L^{ij} \vert^2}{3\vert  C_L^{\rm SM} \vert^2}, \\
  &=&1 + \frac{2 \sum_{i}C_L^{\rm SM} \Delta C_L^{ii} +  \sum_{ij} \vert \Delta C_L^{ij} \vert^2}{3 \vert  C_L^{\rm SM} \vert^2}, 
\end{eqnarray}

\noindent From this expression, we can observe that diagonal leptonic couplings $g^\ell_{\mu\mu}$ and $g^\ell_{\tau\tau}$ contribute to $b\to s\nu_\mu \bar{\nu}_\mu$  (relevant for $b \to s \mu^+\mu^-$ data) and $b\to s\nu_\tau \bar{\nu}_\tau$ (relevant for $b\to c\tau\bar{\nu}_\tau$ data), respectively. In addition, since the neutrino flavor is experimentally unobservable in heavy meson experiments, it is also possible to induce the LFV transitions $b \to s \nu_\mu \bar{\nu}_\tau$ (and $\nu_\tau \bar{\nu}_\mu$) through the off-diagonal coupling $g^\ell_{\mu\tau}$. \medskip

On the experimental side, the Belle experiment in 2017 obtained the following ULs on the branching fractions ${\rm BR}(B \to K \nu\bar{\nu}) < 1.6 \times 10^{-5}$ and ${\rm BR}(B \to K^{\ast} \nu\bar{\nu}) < 2.7 \times 10^{-5}$~\cite{Grygier:2017tzo}, resulting in limits on the ratios, $R^{\nu\bar{\nu}}_{K} < 3.9$ and $R^{\nu\bar{\nu}}_{K^{\ast}} < 2.7$ ($90\%$ C.L.), respectively~\cite{Grygier:2017tzo}. In 2021, based on an inclusive tagging technique, the Belle II experiment reported the bound ${\rm BR}(B^+ \to K^+ \nu\bar{\nu}) < 4.1 \times 10^{-5}$ at $90\%$ C.L.~\cite{Belle-II:2021rof}. A combination of this new result with previous experimental results leads to the weighted average ${\rm BR}(B^+ \to K^+ \nu\bar{\nu}) = (1.1 \pm 0.4) \times 10^{-5}$~\cite{Dattola:2021cmw}. In turn, the ratio $R^{\nu\bar{\nu}}_{K^{+}}$ has been calculated to be, $R^{\nu\bar{\nu}}_{K^{+}} = 2.4 \pm 0.9$~\cite{Browder:2021hbl}. \medskip

The rare $B$ processes $B_s \to \tau^+ \tau^-$ and $B \to K \tau^+ \tau^-$ (induced via $b \to s \tau^+ \tau^-$ transition) are expected to receive significant NP impact. For the leptonic process $B_s \to \tau^+ \tau^-$, the SM branching ratio is shifted by the factor
\begin{equation}
{\rm BR}(B_s \to \tau^+ \tau^-) = {\rm BR}(B_s \to \tau^+ \tau^-)_{\text{SM}} 
 \Bigg| 1+ \dfrac{ \pi}{\sqrt{2} G_F \alpha_{\rm em} V_{tb} V_{ts}^\ast C_{10}^{\rm SM}} \dfrac{g^q_{sb}(g^\ell_{\tau\tau})^\ast}{M_{V}^2} \Bigg|^2,
\end{equation}
\noindent where $C_{10}^{\rm SM} \simeq -4.3$. The strongest experimental bound on its branching ratio has been obtained by the LHCb, ${\rm BR}(B_s \to \tau^+ \tau^-) < 6.8 \times 10^{-3}$ at 95\% confidence level~\cite{Aaij:2017xqt}, while its SM predictions is ${\rm BR}(B_s^0 \to \tau^+ \tau^-)_{\rm SM} =  (7.73 \pm 0.49) \times 10^{-7}$~\cite{Bobeth:2013uxa}. The bound is
 \begin{eqnarray}
\dfrac{\vert g^q_{sb}(g^\ell_{\tau\tau})^\ast \vert}{M_{V}^2}  &<&  0.56 \ {\rm TeV}^{-2}.
\end{eqnarray}



As concerns the semileptonic decay $B \to K \tau^+ \tau^-$, an easy handle numerical formula for the branching ratio (over the whole kinematic range for the lepton pair invariant mass) has been obtained in Ref.~\cite{Cornella:2019hct}, for the case of a singlet vector leptoquark explanation of the $B$ meson anomalies. Since the NP contribution is generated via the same operator, this expression can be easily (but properly) translated to the TVB model, namely 
\begin{equation}
{\rm BR}(B \to K \tau^+ \tau^-) \simeq 1.5\times 10^{-7} + 1.4\times 10^{-3} \Big( \frac{1}{2\sqrt{2}G_F}\Big) \frac{{\rm Re}[g^q_{sb}(g^\ell_{\tau\tau})^\ast]}{M_{V}^2} + 3.5 \Big( \frac{1}{2\sqrt{2}G_F}\Big)^2 \frac{\vert g^q_{sb}(g^\ell_{\tau\tau})^\ast \vert^2}{M_{V}^4}.
\end{equation}

\noindent This decay channel has not been observed so far, and the present reported bound is ${\rm BR}(B \to K \tau^+\tau^-)<2.25\times 10^{-3}$~\cite{PDG2020}. We obtained the following bound
\begin{equation}
 \frac{\vert g^q_{sb}(g^\ell_{\tau\tau})^\ast \vert}{M_{V}^2}  < 0.83\ {\rm TeV}^{-2},
\end{equation}

\noindent that is weaker than the one get from $B_s \to \tau^+ \tau^-$. 

\subsection{$\tau$ decays: $\tau \to 3\mu$, $\tau \to \mu\bar{\nu}_\mu\nu_\tau$, and $\tau \to \mu\phi$}

It is known that the TVB model generates four-lepton operators $(\bar{\mu}\gamma^\alpha P_L \tau)(\bar{\mu}\gamma_\alpha P_L \mu)$ and $(\bar{\mu}\gamma^\alpha P_L \tau)(\bar{\nu}_\tau \gamma_\alpha P_L \nu_\mu)$, thus yielding to tree-level contributions to the leptonic $\tau$ decays, $\tau^- \to \mu^-\mu^+\mu^- \ (\tau \to 3\mu)$ and $\tau^- \to \mu^-\bar{\nu}_\mu\nu_\tau$, respectively~\cite{Bhattacharya:2016mcc,Kumar:2018kmr}. For the LFV decay $\tau \to 3\mu$, the expression for the branching ratio can be written as
\begin{equation}
{\rm BR}(\tau^- \to \mu^-\mu^+\mu^-) = \frac{m_\tau^5}{1536\pi^3 \Gamma_\tau} \frac{\vert g^\ell_{\mu\mu} g^\ell_{\mu\tau}\vert^2}{M_V^4},
\end{equation}

\noindent where $\Gamma_\tau$ is the total decay width of the $\tau$ lepton. The current experimental UL obtained by Belle (at 90\% CL) is ${\rm BR}(\tau^- \to \mu^-\mu^+\mu^-) < 2.1 \times 10^{-8}$~\cite{PDG2020}. This corresponds to
\begin{equation}
\frac{\vert g^\ell_{\mu\mu} g^\ell_{\mu\tau}\vert}{M_V^2}  < 1.13\times 10^{-2} \ {\rm TeV}^{-2}.
\end{equation}


The leptonic decay $\tau^- \to \mu^-\bar{\nu}_\mu\nu_\tau$ is a lepton flavor conserving and SM allowed process that receives tree-level contribution from both $W^\prime$ (via lepton flavor conserving couplings) and $Z^\prime$ (via LFV couplings) bosons~\cite{Kumar:2018kmr}. The branching ratio is given by~\cite{Kumar:2018kmr}
\begin{equation}
{\rm BR}(\tau \to \mu \bar{\nu}_\mu\nu_\tau) = {\rm BR}(\tau \to \mu \bar{\nu}_\mu\nu_\tau)_{\rm SM} \bigg(\bigg\vert 1+ \dfrac{1}{2\sqrt{2}G_F M_V^2} (2 g^\ell_{\mu\mu}g^\ell_{\tau\tau} - \vert g^\ell_{\mu\tau}\vert^2) \bigg\vert^2 + \bigg\vert \dfrac{1}{2\sqrt{2}G_F M_V^2} \vert g^\ell_{\mu\tau}\vert^2 \bigg\vert^2 \bigg),
\end{equation}

\noindent where ${\rm BR}(\tau \to \mu \bar{\nu}_\mu\nu_\tau)_{\rm SM} = (17.29 \pm 0.03) \%$~\cite{Pich:2013lsa}. The $Z^\prime$ boson can also generates one-loop corrections, which can be safely ignored. This value has to be compared with the experimental value reported by PDG ${\rm BR}(\tau \to \mu \bar{\nu}_\mu\nu_\tau) = (17.39 \pm 0.04) \%$~\cite{PDG2020}. \medskip 

Finally, the branching ratio of the LFV hadronic $\tau$ decay $\tau \to \mu \phi$ ($\tau \to \mu s\bar{s}$ transition), can be expressed as~\cite{Bhattacharya:2016mcc}
\begin{equation}
{\rm BR}(\tau^- \to \mu^-\phi) =  \frac{f_\phi^2 m_\tau^3}{128\pi \Gamma_\tau} \Big(1+2 \frac{m_\phi^2}{m_\tau^2} \Big) \Big(1- \frac{m_\phi^2}{m_\tau^2} \Big)^2 \frac{\vert g^\ell_{\mu\tau} g^q_{ss}\vert^2}{M_V^4},
\end{equation}

\noindent where $m_\phi$ and  $f_\phi = (238 \pm 3)$ MeV~\cite{Kumar:2018kmr} are the $\phi$ meson mass and  decay constant, respectively. Currently, the UL reported by Belle on the branching ratio is  ${\rm BR}(\tau^- \to \mu^-\phi) < 2.3\times 10^{-8}$~\cite{Belle:2023ziz}. The current UL produces the bound
\begin{equation}
\frac{\vert g^\ell_{\mu\tau} g^q_{ss}\vert}{M_V^2}  < 9.4 \times 10^{-3}  \ {\rm TeV}^{-3}.
\end{equation}

\noindent Since the $D^0 -\bar{D}^0$ mixing imposes that $\vert g^q_{ss}\vert/M_V \leq 3.3 \times 10^{-3}  \ {\rm TeV^{-1}}$ (see Sec.~\ref{mixing}) the constraint from $\tau \to \mu \phi$ can be easily fulfilled. We will not take into account this LFV process in further TVB model analysis.

\subsection{LHC bounds} 
\label{LHC}

LHC constraints are always important for models with  non-zero $Z^{\prime}$ couplings to the SM particles~\cite{Langacker:2008yv}. {In particular, in our study it will set important constraints on the parametric space conformed by the TVB couplings $(g^q_{bb},g^\ell_{\mu\mu})$ and $(g^q_{bb},g^\ell_{\tau\tau})$}. We consider the ATLAS search for high-mass dilepton resonances in the mass range of 250~GeV to 6~TeV,
in proton-proton collisions at a center-of-mass energy of $\sqrt{s}=13$ TeV during Run 2 of the LHC with an integrated luminosity of 139~fb$^{-1}$~\cite{ATLAS:2019erb}~(recently, the CMS collaboration has also reported constraints for similar luminosities~\cite{CMS:2019tbu}, basically identical to ATLAS~\cite{ATLAS:2019erb}), and the  data from searches of $Z^{\prime}$ bosons decaying to tau pairs with an integrated luminosity of 36.1 fb$^{-1}$ from proton-proton collisions at $\sqrt{s}= 13$ TeV~\cite{ATLAS:2017eiz}. There are also searches for high-mass resonances in the monolepton channels ($pp \to \ell\nu$) carried out by the ATLAS and CMS~\cite{ATLAS:2019lsy,ATLASmonotau,CMS:2022ncp}. However, they provide weaker bounds than those obtained from dilepton searches, and we will not take them into account.

We obtain for benchmark mass value $M_{V}=1$~TeV the lower limit on the parameter space from the intersection of the 95$\%$CL upper limit on the cross-section from the ATLAS experiment~\cite{ATLAS:2019erb,ATLAS:2017eiz} with the theoretical cross-section given in Ref.~\cite{Erler:2011ud}. Lower limits above $4.5$~TeV apply to models with couplings to the first family, which it is not our case. 
The strongest restrictions come from $Z^{\prime}$ production processes in the $b\bar{b}$ annihilation and the subsequent $Z^{\prime}$ decay into muons ($\mu^+\mu^-$) and taus ($\tau^+\tau^-$). Further details are shown in Refs.~\cite{Erler:2011ud,Salazar:2015gxa,Benavides:2018fzm}. Let us remark that within the TVB framework is also possible to consider the annihilation between quarks with different flavors (namely, $g^q_{bs}$), however, we anticipate that according to our phenomenological analysis in Sec.~\ref{analysis}  this coupling is very small; therefore, we only consider production processes without flavor changing neutral currents. In the next section we will show that the TVB parameter space is limited by LHC constraints to regions where the couplings of the leptons or the quarks are close to zero, excluding  the regions preferred by the $B$ meson anomalies and low-energy flavor observables. \medskip 

\section{Analysis on the TVB parametric space} \label{analysis}

In this section we present the parametric space analysis of the TVB model addressing a simultaneous explanation of the $b\to s \mu^+\mu^-$ and $b \to c \tau\bar{\nu}_\tau$ data. We define the pull for the $i$-th observable as
\begin{equation} \label{pull}
{\rm pull}_i = \frac{\mathcal{O}^{\rm exp}_i - \mathcal{O}^{\rm th}_i}{\Delta \mathcal{O}_i},
\end{equation}

\noindent where $\mathcal{O}^\text{exp}_i$ is the experimental measurement, $\mathcal{O}^\text{th}_i \equiv \mathcal{O}^\text{th}_i(g^q_{bs},g^q_{bb},g^\ell_{\mu\mu},g^\ell_{\tau\tau},g^\ell_{\mu\tau})$ is the theoretical prediction that include the NP contributions, and $\Delta \mathcal{O}_i =  ((\sigma^{\rm exp}_i)^2 + (\sigma^{\rm th}_i)^2)^{1/2}$ corresponds to the combined experimental and theoretical uncertainties. By means of the pull, we can compare the fitted values of each observable to their measured values. The $\chi^2$ function is written as the sum of squared pulls, i.e.,
\begin{equation}
\chi^2 = \sum_{i}^{N_{\rm obs}} ({\rm pull}_i)^2,
\end{equation}

\noindent where the sum extends over the number of observables $(N_{\rm obs})$ to be fitted. 
Our phenomenological analysis is based on the flavor observables presented in the previous Sec.~\ref{Obs}. This all data set includes: $b \to c \tau \bar{\nu}_\tau$ and $b \to s \mu^+\mu^-$ data, bottomonium ratios $R_{\Upsilon(nS)}$, LFV decays ($B^+ \to K^+ \mu^\pm \tau^\mp$, $B^0 \to K^{\ast 0}\mu^\pm\tau^\mp$, $B_s \to \mu^\pm \tau^\mp$, $\Upsilon(nS) \to \mu^\pm \tau^\mp$), rare $B$ decays ($B \to K^{(\ast)} \nu\bar{\nu}, B \to K \tau^+ \tau^-, B_s \to \tau^+ \tau^-$), $\tau$ decays ($\tau \to 3\mu$, $\tau \to \mu\bar{\nu}_\mu\nu_\tau$), $\Delta F = 2$ processes, and neutrino trident production. 
We will study the impact of the most recent LHCb measurements on the ratios $R(D^{(\ast)})$~\cite{LHCb2022,LHCb:2023zxo,LHCb2023}, allowing us to present an updated status of the TVB model as an explanation to the $B$ meson anomalies. For such a purpose, we will consider in our analysis the following three different sets of observables,
\begin{itemize}
\item All data with $R(D)_{\rm LHCb22}$ + $R(D^\ast)_{\rm LHCb23}$,
\item All data with $R(D^{(\ast)})_{\rm LHCb22}$,
\item All data with $R(D^{(\ast)})_{\rm HFLAV23}$.
\end{itemize}

\noindent All these three sets contain a total number of observables $N_{\rm obs} = 31$ and five free TVB parameters ($g^q_{bs}$, $g^q_{bb}$, $g^\ell_{\mu\mu}$, $g^\ell_{\tau\tau}$, $g^\ell_{\mu\tau}$) to be fitted. The heavy TVB mass will be fixed to the benchmark value $M_{V} = 1 \ {\rm TeV}$. Therefore, the number of degrees of freedom is $N_{\rm dof} = 26$. \medskip


\begin{table}[t]
\centering
\renewcommand{\arraystretch}{1.4}
\renewcommand{\arrayrulewidth}{0.8pt}
\caption{Best-fit point values and $1\sigma$ intervals of the five TVB couplings $(g^q_{bs},g^q_{bb},g^\ell_{\mu\mu},g^\ell_{\tau\tau},g^\ell_{\mu\tau})$ for the three different sets of observables and a benchmark mass value of $M_{V} = 1 \ {\rm TeV}$.} \label{fit}
\begin{tabular}{ccc}
\hline\hline
TVB couplings & Best-fit point  & $1\sigma$ intervals \\                                                                                                                        
\hline
\multicolumn{3}{c}{All data with $R(D)_{\rm LHCb22}$ + $R(D^\ast)_{\rm LHCb23}$ : $\chi^2_{\rm min} /N_{\rm dof} = 0.63$, $p$-value $= 93.7 \%$} \\
\hline
$g^q_{bs}$  & $-2.3\times 10^{-3}$ & $[-3.2,-1.6]\times 10^{-3}$ \\ 
$g^q_{bb}$  & 0.73 & $[0.28,1.72]$ \\
$g^\ell_{\mu\mu}$  & 0.20 & $[0.072,0.131]$ \\
$g^\ell_{\tau\tau}$  & 0.49 & $[0.27,0.71]$ \\
$g^\ell_{\mu\tau}$  & $\sim 0$ & $[-0.11,0.11]$ \\
\hline
\multicolumn{3}{c}{All data with $R(D^{(\ast)})_{\rm LHCb22}$ : $\chi^2_{\rm min} /N_{\rm dof} = 0.62$, $p$-value $= 93.1 \%$} \\
\hline
$g^q_{bs}$  & $-3.2\times 10^{-3}$ & $[-4.4,-2.1]\times 10^{-3}$  \\ 
$g^q_{bb}$  & 1.50 & $[0.74,2.24]$ \\
$g^\ell_{\mu\mu}$  & 0.074 & $[0.052,0.095]$ \\
$g^\ell_{\tau\tau}$  & 0.70 & $[0.45,0.94]$ \\
$g^\ell_{\mu\tau}$  & $\sim 0$ & $[-0.15,0.15]$ \\
\hline
\multicolumn{3}{c}{All data with $R(D^{(\ast)})_{\rm HFLAV23}$ : $\chi^2_{\rm min} /N_{\rm dof} = 0.59$, $p$-value $= 95.2 \%$} \\
\hline
$g^q_{bs}$  & $-3.2\times 10^{-3}$ & $[-4.4,-2.1]\times 10^{-3}$ \\ 
$g^q_{bb}$  & 1.52 & $[1.09,1.94]$ \\
$g^\ell_{\mu\mu}$  & 0.073 & $[0.052,0.095]$ \\
$g^\ell_{\tau\tau}$  & 0.70 & $[0.53,0.88]$ \\
$g^\ell_{\mu\tau}$  & $\sim 0$ & $[-0.14,0.14]$ \\
\hline\hline
\end{tabular}
\end{table} 

For the three sets of observables we find the best-fit point values by minimizing the $\chi^2$ function ($\chi^2_{\rm min}$). In Table~\ref{fit} we report our results of the best-fit point values and $1\sigma$ intervals of TVB couplings. For each fit we also present in Table~\ref{fit} the values of $\chi^2_{\rm min} /N_{\rm dof}$ and its corresponding $p$-value to evaluate the fit-quality.
In general, it is found that the three sets of observables provide an excellent fit of the data.
In the quark sector, the TVB model requires small $g^q_{bs}$ coupling, $\vert g^q_{bs} \vert \sim \mathcal{O}(10^{-3})$, and opposite sign to $g^\ell_{\mu\mu}$ to be consistent with $b\to s \mu^+\mu^-$ data ($C_9^{\mu\mu}= - C_{10}^{\mu\mu}$ solution) and $B_s -\bar{B}_s$ mixing. On the other hand, large values for the bottom-bottom coupling $g^q_{bb}\sim \mathcal{O}(1)$ are preferred. As for the leptonic couplings, it is found that the lepton flavor conserving ones have a similar size $g^\ell_{\mu\mu}  \approx  g^\ell_{\tau\tau} \sim \mathcal{O}(10^{-1})$ for All data with $R(D)_{\rm LHCb22}$ + $R(D^\ast)_{\rm LHCb23}$, suggesting non-hierarchy pattern. While for All data with $R(D^{(\ast)})_{\rm LHCb23}$ (with $R(D^{(\ast)})_{\rm HFLAV23}$), these couplings exhibit a hierarchy $g^\ell_{\tau\tau} > g^\ell_{\mu\mu}$. 
As LFV coupling concerns, the obtained best-fit point values on $g^\ell_{\mu\tau}$ are negligible. Thus, the TVB model do not lead to appreciable LFV effects. 
Last but no least, we also probe higher mass values ($M_{V} > 1 \ {\rm TeV}$). We obtain that in order to avoid large values on $g^q_{bb}$ coupling $(\sim \sqrt{4\pi})$, that would put the perturbativity of the model into question, the TVB mass can be as large as $M_V \sim 2$ TeV. \medskip

\begin{figure*}[!t]
(a) All data with $R(D)_{\rm LHCb22}$ + $R(D^\ast)_{\rm LHCb23}$ \medskip

\includegraphics[scale=0.195]{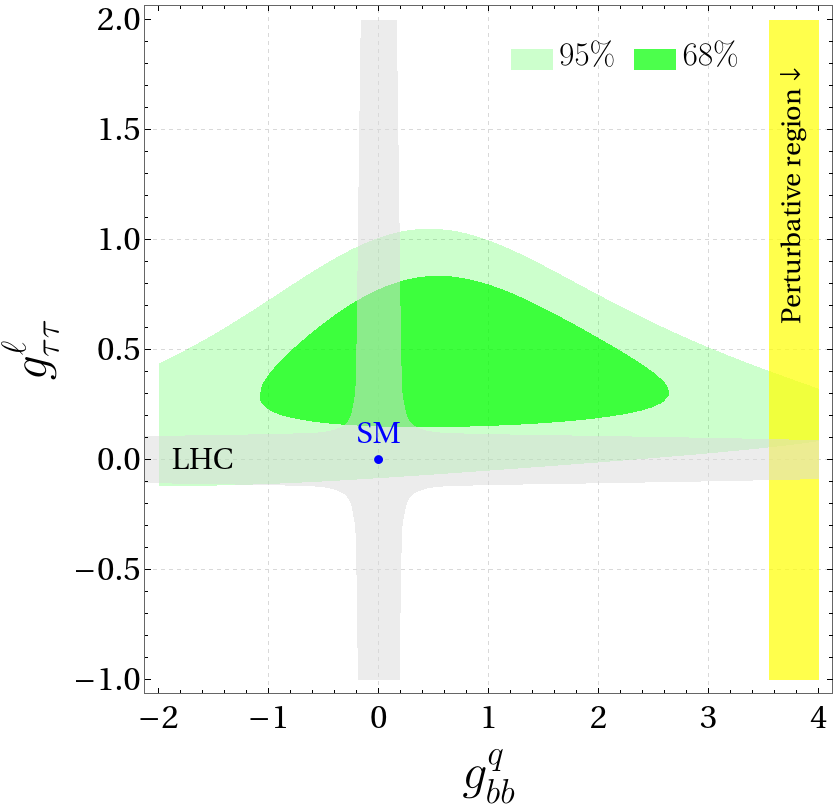} \ 
\includegraphics[scale=0.195]{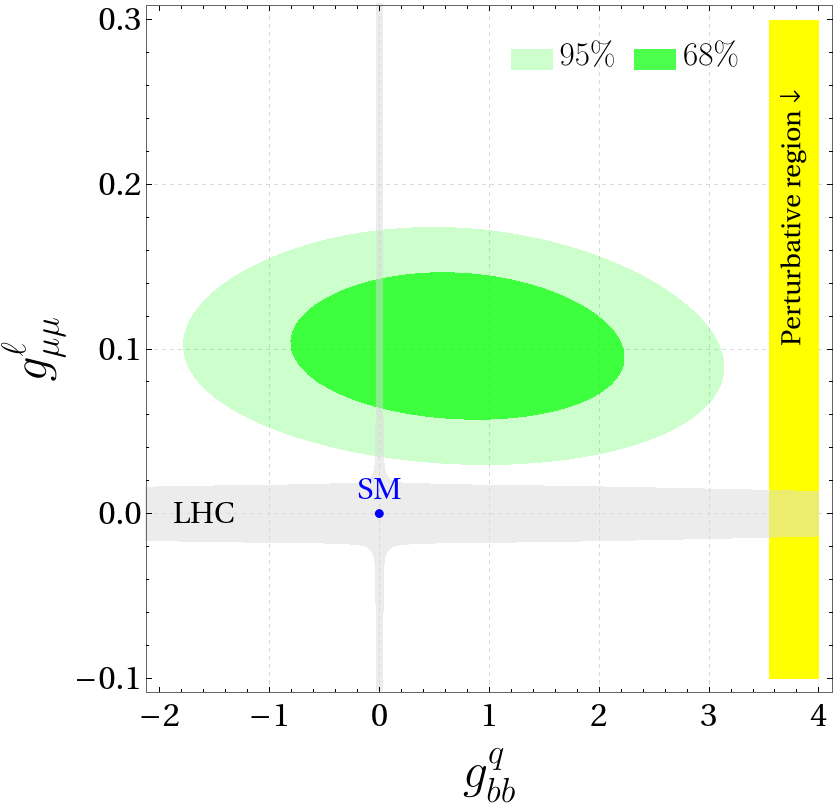} \ 
\includegraphics[scale=0.2]{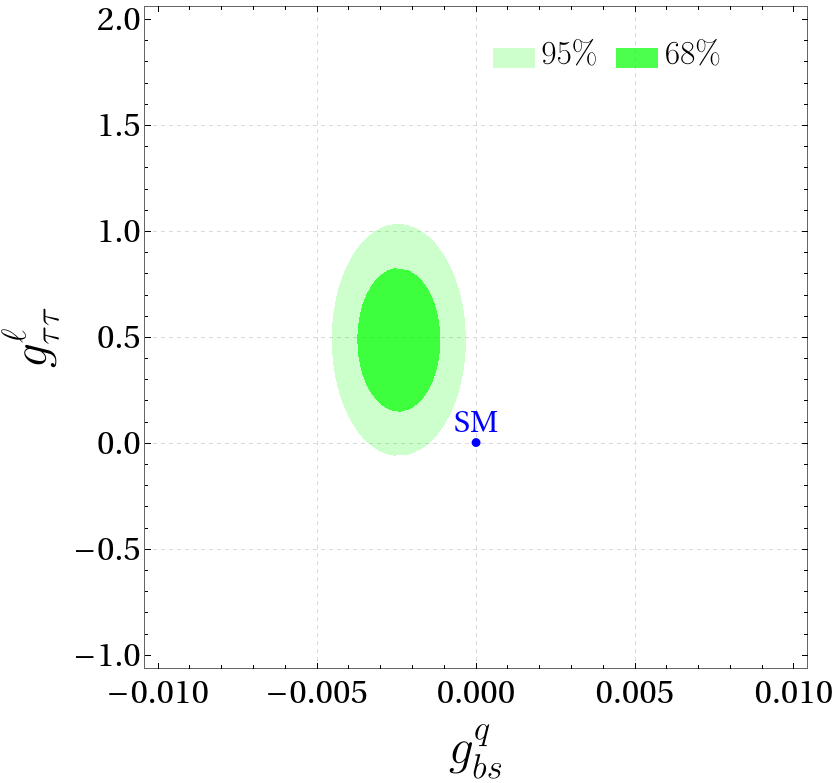} \ 
\includegraphics[scale=0.2]{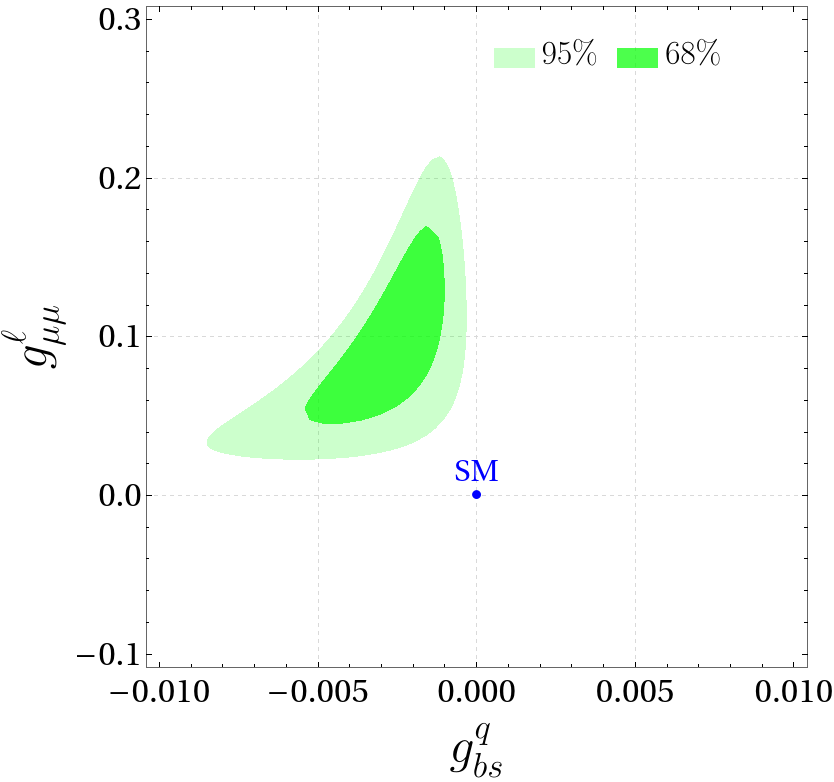}\medskip

(b) All data with $R(D^{(\ast)})_{\rm LHCb23}$ \medskip

\includegraphics[scale=0.195]{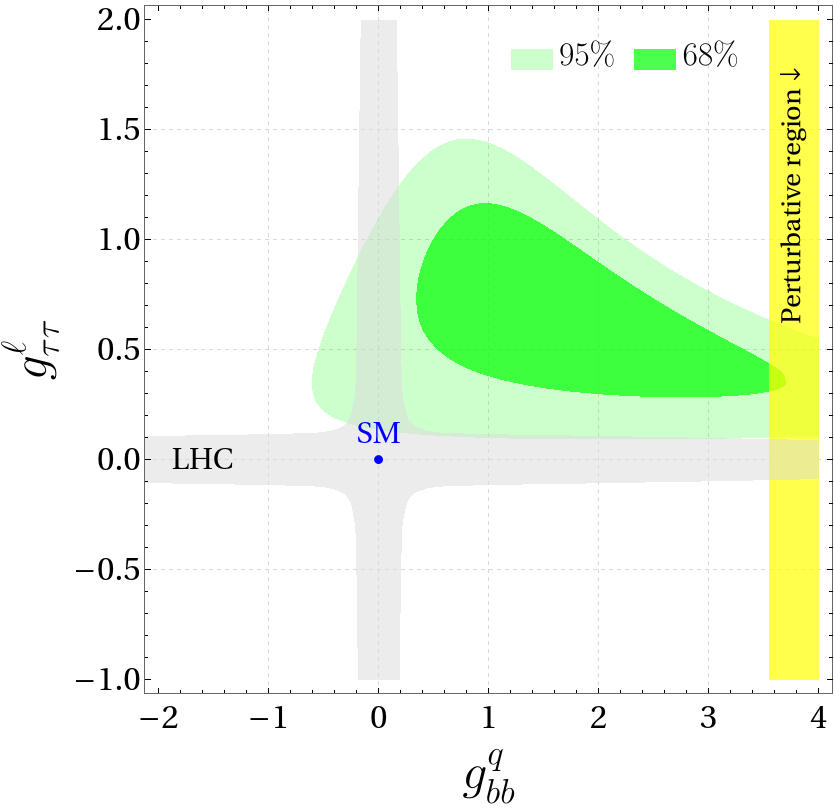} \
\includegraphics[scale=0.195]{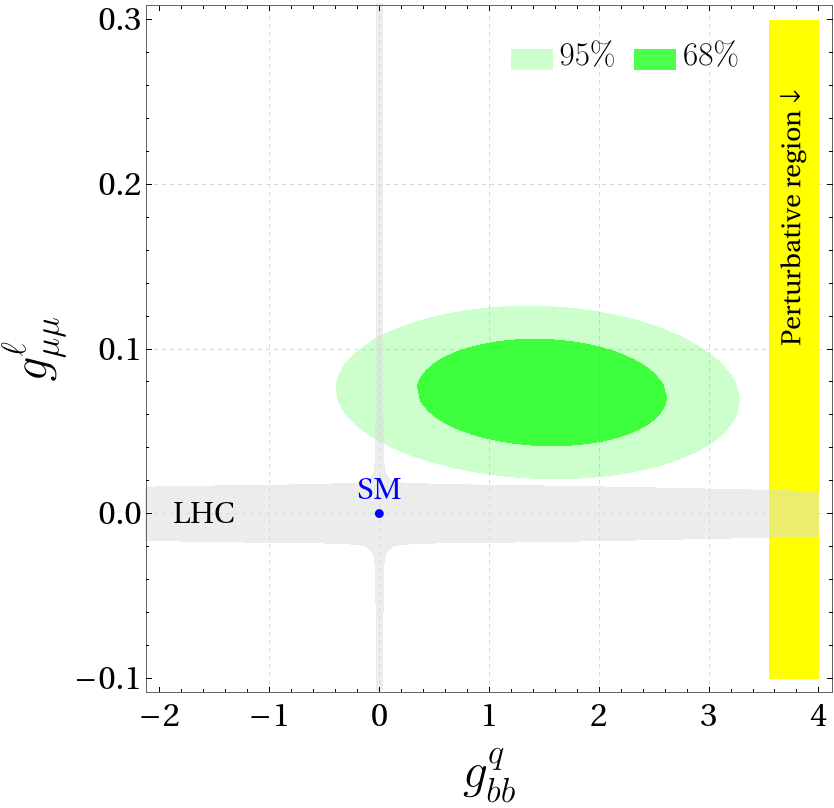}  \
\includegraphics[scale=0.2]{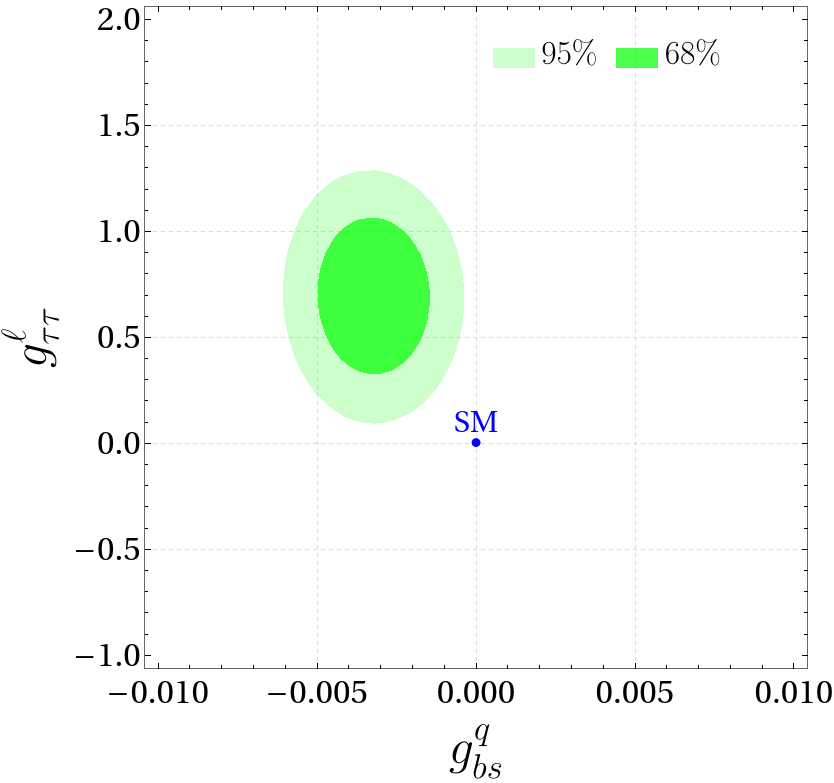} \ 
\includegraphics[scale=0.2]{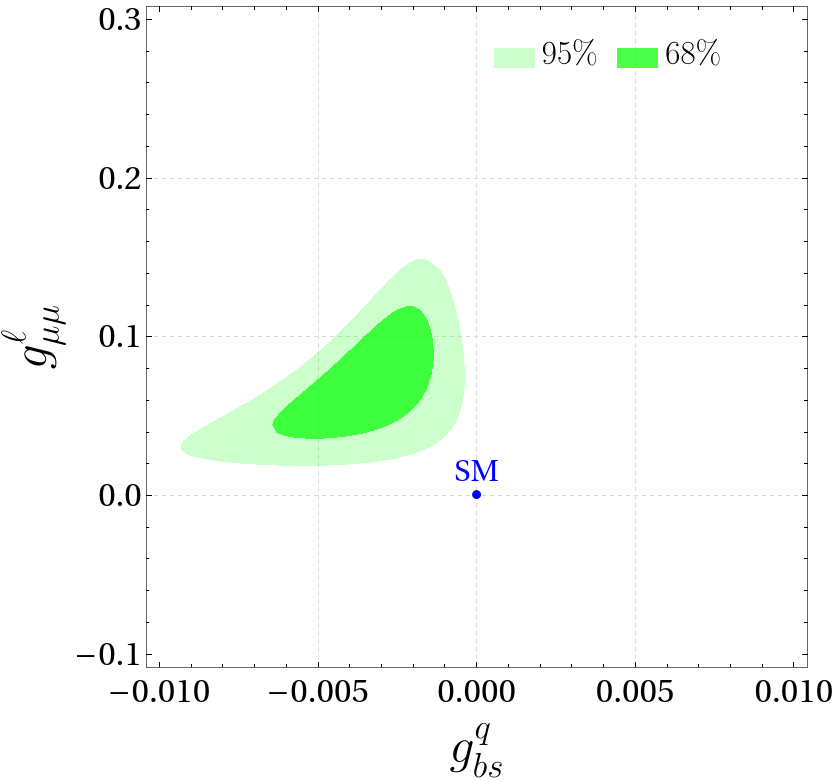}\medskip

(c) All data with $R(D^{(\ast)})_{\rm HFLAV23}$ \medskip

\includegraphics[scale=0.195]{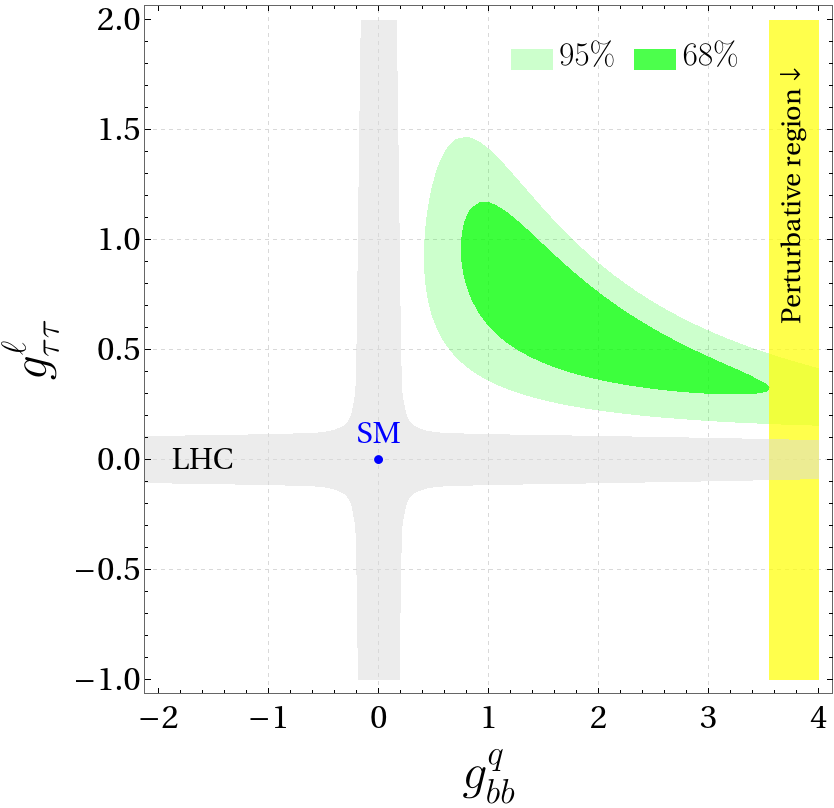} \
\includegraphics[scale=0.195]{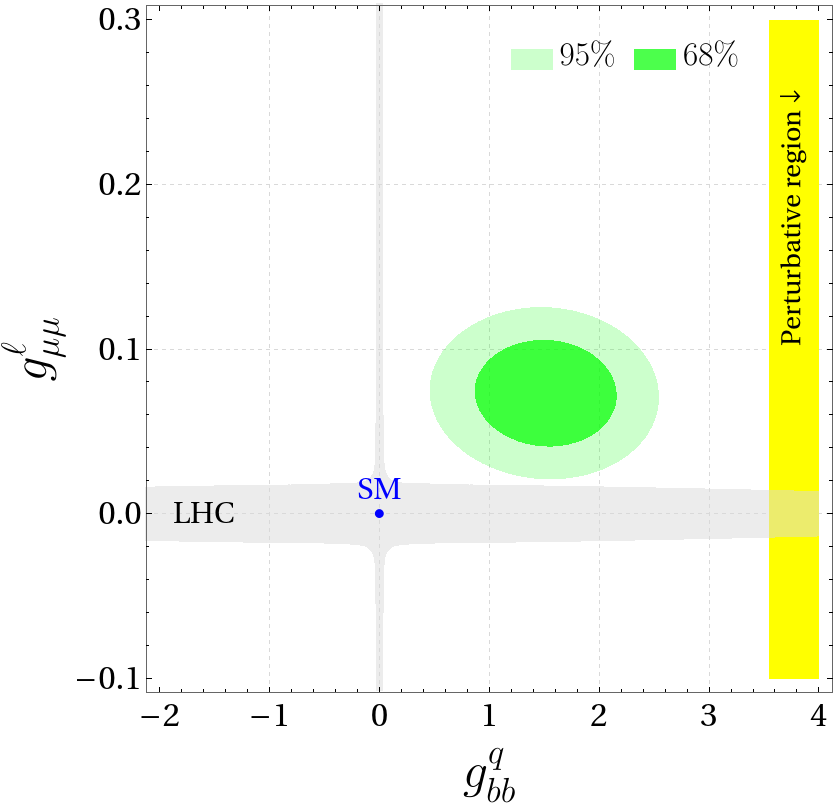} \
\includegraphics[scale=0.2]{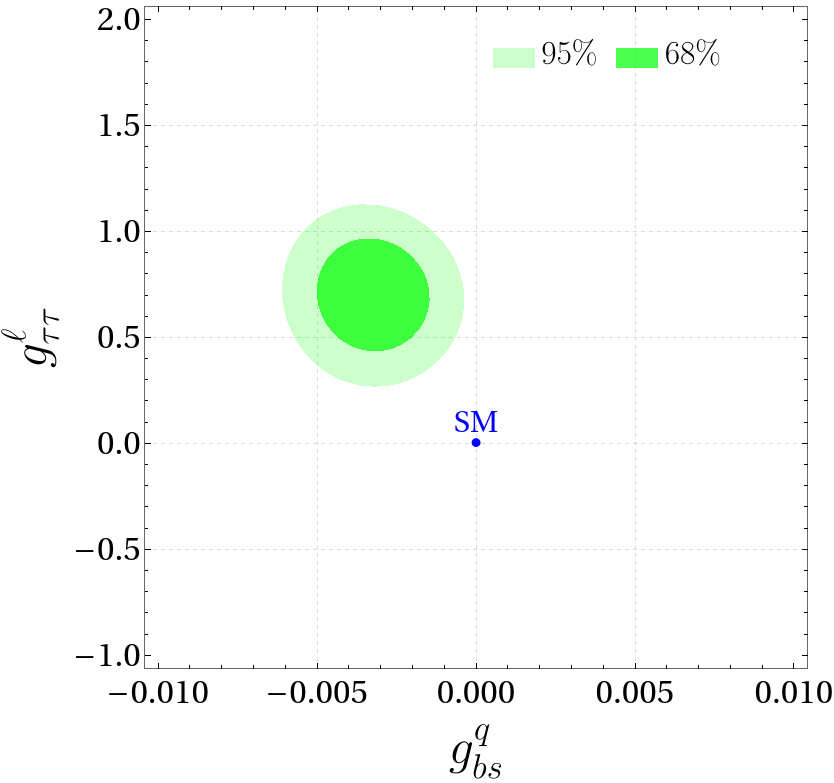} \ 
\includegraphics[scale=0.2]{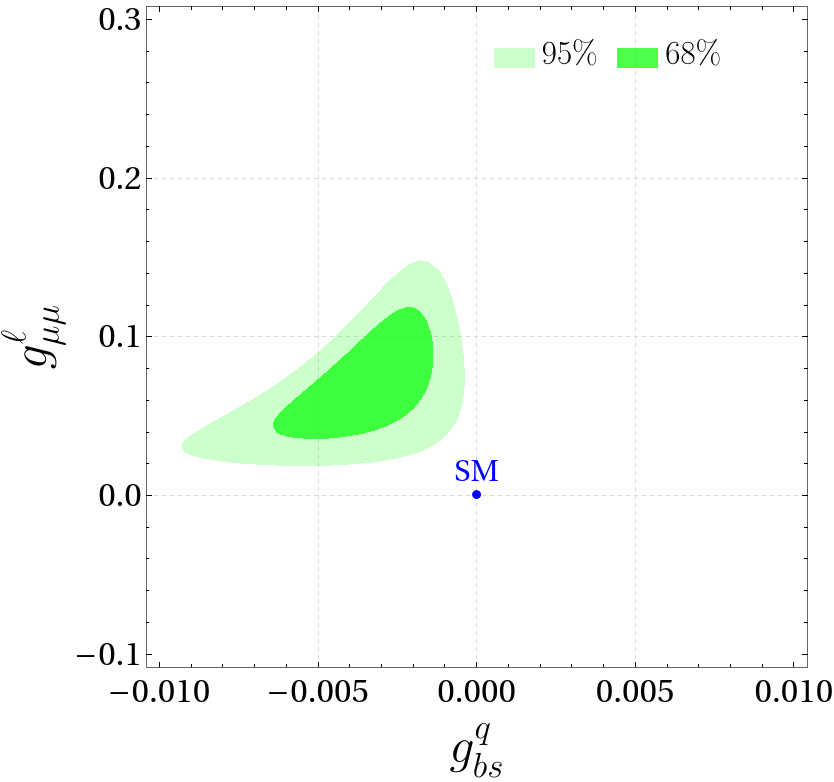}

\caption{\small $68\%$ (green) and $95\%$ (light-green) CL allowed regions for the most relevant 2D parametric space of (a) All data with $R(D)_{\rm LHCb22}$ + $R(D^\ast)_{\rm LHCb23}$, (b) All data with $R(D^{(\ast)})_{\rm LHCb22}$, and (c) All data with $R(D^{(\ast)})_{\rm HFLAV23}$, respectively, for $M_{V} = 1 \ {\rm TeV}$. In each plot we are marginalizing over the rest of the parameters. The SM value is represented by the blue dot. The light-gray region corresponds to LHC bounds at the $95\%$ CL. Perturbative region ($g^q_{bb} \geq \sqrt{4\pi})$) is represented by yellow color.}
\label{PS}
\end{figure*}

In Fig.~\ref{PS}, we show the allowed regions of the most relevant two-dimension (2D) parametric space of (a) All data with $R(D)_{\rm LHCb22}$ + $R(D^\ast)_{\rm LHCb23}$, (b) All data with $R(D^{(\ast)})_{\rm LHCb22}$, and (c) All data with $R(D^{(\ast)})_{\rm HFLAV23}$, respectively, for a benchmark TVB mass $M_{V} = 1 \ {\rm TeV}$. The $68\%$ and $95\%$ CL regions are shown in green and light-green colors, respectively. In each plot we are marginalizing over the rest of the parameters.  
Furthermore, we include the LHC bounds (light-gray regions) obtained from searches of high-mass dilepton (dimuon and ditau) resonances at the ATLAS experiment~\cite{ATLAS:2019erb,ATLAS:2017eiz}, as discussed in Sec.~\ref{LHC}. For completeness, the perturbative region ($g^q_{bb} \geq \sqrt{4\pi})$) is represented by yellow color.
It is observed in the planes ($g^q_{bb},g^\ell_{\tau\tau}$) and ($g^q_{bb},g^\ell_{\mu\mu}$) for All data with $R(D^{(\ast)})_{\rm HFLAV23}$ that the TVB model is seems to be strongly ruled out by the LHC bounds. However, for All data with $R(D)_{\rm LHCb22}$ + $R(D^\ast)_{\rm LHCb23}$ (and with $R(D^{(\ast)})_{\rm LHCb22}$) that include the very recent LHCb measurements~\cite{LHCb2022,LHCb:2023zxo,LHCb2023}, the TVB model can provide a combined explanation of the $b \to c \tau \bar{\nu}_\tau$ and $b \to s \mu^+\mu^-$ anomalies, in consistency with LHC bounds. Our analysis shows that given the current experimental situation, particularly with LHCb, it is premature to exclude the TVB model to addressing the $B$ meson anomalies. Future improvements and new measurements on $b \to c \tau\bar{\nu}_\tau$ data at the Belle II and LHCb experiments will be a matter of importance to test the TVB model.\medskip

We close by mentioning that an analysis of the TVB model was previously reported by Kumar, London, and Watanabe (KLW) by implementing the 2018 $b \to c \tau \bar{\nu}_\tau$ and $b \to s \mu^+\mu^-$ data~\cite{Kumar:2018kmr}. KLW found that the TVB model is excluded as a possible explanation of the $B$ meson anomalies due to the bound from LHC dimuon search (3.2 fb${}^{-1}$)~\cite{Kumar:2018kmr}. Such a result is in agreement with ours for All data with $R(D^{(\ast)})_{\rm HFLAV23}$ and considering recent LHC dimuon (139 fb${}^{-1}$) and ditau (36.1 fb${}^{-1}$) searches. Unlike to KLW analysis, we have incorporated several new observables and considered the recent available experimental measurements and ULs. Thus, our present study extends, complements, and update the previous analysis performed by KLW. We also extend the recent analysis~\cite{Garcia-Duque:2021qmg} where only the charged-current $b \to c \tau\bar{\nu}_\tau$ anomaly was addressed within this framework. 

\subsection{Implications to some flavor parametrizations}

As a final step in our analysis, we will explore the implications to our previous phenomenological analysis on TVB model to some flavor parametrizations that have been already studied in the literature. For this we consider scenarios in which the transformations involve only the second and third generations~\cite{Bhattacharya:2016mcc,Calibbi:2015kma}, as it was previously discussed in Sec.~\ref{TVB}, we found that the equivalence in the quark sector is Eq.~\eqref{gtogquark}, while for the leptonic sector we have Eq.~\eqref{gtoglepton}.
Taking into account the $1\sigma$ range solutions of TVB couplings obtained in Table~\ref{fit} (for the three sets of data), we get, in general, a large coupling $g_{2}^{q} \sim \mathcal{O}(1)$ and a very small mixing angle $\vert \theta_D \vert \sim 10^{-3}$. Such a small mixing angle ($\vert \theta_D \vert \ll V_{cb}$) result is still in agreement with previous analysis~\cite{Bhattacharya:2016mcc,Calibbi:2015kma}. On the contrary, in the leptonic sector, we obtained that because of $1\sigma$ range of the LFV coupling $g_{\mu\tau}^\ell$ it is not possible to find a physical solution to the mixing angle $\theta_L$. As additional probe, we have performed a global fit to the current $b \to s \mu^+\mu^-$ and $b \to c \tau \bar{\nu}_\tau$ data, and the most relevant flavor observables, with $(g_{2}^{q}, g_{2}^{\ell},\theta_D,\theta_L)$ as free parameters. For a fixed mass value $M_V = 1 \ {\rm TeV}$, we obtained a very poor fit ($\chi^2_{\rm min} /N_{\rm dof} \gg 1$), concluding that this kind of flavor setup is not viable within the TVB model.

\section{Conclusions} \label{Conclusion}

We have presented an updated view of the TVB model as a simultaneous explanation of the $B$ meson anomalies ($b \to c \tau \bar{\nu}_\tau$ and $b \to s \mu^+\mu^-$ data). We performed a global fit of the TVB parameter space with the most recent 2022 and 2023 data, including the LHCb measurements on the charged-current LFU ratios $R(D^{(\ast)})$ and $R(\Lambda_c)$. As concerns $b \to s \mu^+\mu^-$ data, we taken into account the $C^{bs\mu\mu}_{9} = - C^{bs\mu\mu}_{10}$ solution from global fit analysis including the recent results on $R_{K^{(\ast)}}$ by LHCb and ${\rm BR}(B_s \to \mu^+\mu^-)$ by CMS. We have also 
included all relevant flavor observables such as $B_s - \bar{B}_s$ mixing, neutrino trident production, LFV decays ($B \to K^{(\ast)} \mu^\pm \tau^\mp$, $B_s \to \mu^\pm \tau^\mp$, $\tau \to \mu\phi$, $\Upsilon(nS) \to \mu^\pm \tau^\mp$), rare $B$ decays ($B \to K^{(\ast)} \nu\bar{\nu}, B \to K \tau^+ \tau^-, B_s \to \tau^+ \tau^-$), and bottomonium LFU ratios. We have confronted the allowed paramater space with the LHC bounds from searches of high-mass dilepton resonances at the ATLAS experiment.

Our analysis has shown that for a heavy TVB mass of 1 TeV and using all data along with world averages values on $R(D^{(\ast)})$ reported by HFLAV, the TVB model can accommodate the $b \to c \tau \bar{\nu}_\tau$ and $b \to s \mu^+\mu^-$ anomalies (in consistency with other flavor observables), but it seems to be strongly disfavoured by the LHC bounds. However, we obtained a different situation when all data are combined with the very recent LHCb measurements on $R(D^{(\ast)})$. The the $B$ meson anomalies can be addressed within the TVB model in consistency with LHC constraints. We concluded that new and improved $b \to c \tau\bar{\nu}_\tau$ data by LHCb and Belle II will be required to really establish the viability of the TVB model.

We have also studied the consequences of our analysis of the TVB model to flavor parametrizations in which the transformations involve only the second and third generations. We obtained that such a flavor ansatz is not viable within the TVB model.


\acknowledgments

J. H. M. is grateful to Vicerrectoría de Investigación-Creación of Universidad del Tolima for financial support of Project No. 290130517. E. R. acknowledges financial support from the “Vicerrectoría de Investigaciones e Interacción Social VIIS de la Universidad de Nariño,” Projects No. 1928 and No. 2172. We are grateful to Hector Gisbert for his comments on LFV effects in the dineutrino channels $B \to K^{(\ast)} \nu\bar{\nu}$.

\FloatBarrier


\begin{thebibliography}{99}

\bibitem{London:2021lfn}
D.~London and J.~Matias, $B$ Flavour Anomalies: 2021 Theoretical Status Report,
Ann. Rev. Nucl. Part. Sci. \textbf{72}, 37-68 (2022)
[arXiv:2110.13270 [hep-ph]].

\bibitem{Albrecht:2021tul}
J.~Albrecht, D.~van Dyk and C.~Langenbruch, Flavour anomalies in heavy quark decays, Prog. Part. Nucl. Phys. \textbf{120}, 103885 (2021)
[arXiv:2107.04822 [hep-ex]].

\bibitem{Bifani:2018zmi}
S.~Bifani, S.~Descotes-Genon, A.~Romero Vidal and M.~H.~Schune, Review of Lepton Universality tests in $B$ decays, J. Phys. G \textbf{46}, no.2, 023001 (2019)
[arXiv:1809.06229 [hep-ex]].


\bibitem{Aaij:2014ora}
R.~Aaij \textit{et al.} [LHCb], Test of lepton universality using $B^{+}\rightarrow K^{+}\ell^{+}\ell^{-}$ decays, 
Phys. Rev. Lett. \textbf{113}, 151601 (2014)
[arXiv:1406.6482 [hep-ex]].

\bibitem{Aaij:2021vac}
R.~Aaij \textit{et al.} [LHCb], Test of lepton universality in beauty-quark decays, [arXiv:2103.11769 [hep-ex]].

\bibitem{LHCb:2021lvy}
R.~Aaij \textit{et al.} [LHCb], Tests of lepton universality using $B^0\to K^0_S \ell^+ \ell^-$ and $B^+\to K^{*+} \ell^+ \ell^-$ decays, [arXiv:2110.09501 [hep-ex]].

\bibitem{Aaij:2019wad}
R.~Aaij \textit{et al.} [LHCb], Search for lepton-universality violation in $B^+\to K^+\ell^+\ell^-$ decays,
Phys. Rev. Lett. \textbf{122}, no.19, 191801 (2019)
[arXiv:1903.09252 [hep-ex]].

\bibitem{Aaij:2017vbb}
R.~Aaij \textit{et al.} [LHCb], Test of lepton universality with $B^{0} \rightarrow K^{*0}\ell^{+}\ell^{-}$ decays, JHEP \textbf{08}, 055 (2017)
[arXiv:1705.05802 [hep-ex]].

\bibitem{LHCb:2022qnv}
 [LHCb], Test of lepton universality in $b \rightarrow s \ell^+ \ell^-$ decays, [arXiv:2212.09152 [hep-ex]].

\bibitem{LHCb:2022zom}
 [LHCb], Measurement of lepton universality parameters in $B^+\to K^+\ell^+\ell^-$ and $B^0\to K^{*0}\ell^+\ell^-$ decays, [arXiv:2212.09153 [hep-ex]].


\bibitem{CMS:2022mgd}
 [CMS], Measurement of the B$^0_\mathrm{S}$$\to$$\mu^+\mu^-$ decay properties and search for the B$^0$$\to$$\mu^+\mu^-$ decay in proton-proton collisions at $\sqrt{s}$ = 13 TeV,
[arXiv:2212.10311 [hep-ex]].

\bibitem{Aaij:2013qta}
R.~Aaij \textit{et al.} [LHCb], Measurement of Form-Factor-Independent Observables in the Decay $B^{0} \to K^{*0} \mu^+ \mu^-$, Phys. Rev. Lett. \textbf{111}, 191801 (2013)
[arXiv:1308.1707 [hep-ex]].

\bibitem{Aaij:2015oid}
R.~Aaij \textit{et al.} [LHCb], Angular analysis of the $B^{0} \to K^{*0} \mu^{+} \mu^{-}$ decay using 3 fb$^{-1}$ of integrated luminosity, JHEP \textbf{02}, 104 (2016)
[arXiv:1512.04442 [hep-ex]].

\bibitem{Aaij:2020nrf}
R.~Aaij \textit{et al.} [LHCb], Measurement of $CP$-Averaged Observables in the $B^{0}\rightarrow K^{*0}\mu^{+}\mu^{-}$ Decay, Phys. Rev. Lett. \textbf{125}, no.1, 011802 (2020)
[arXiv:2003.04831 [hep-ex]].

\bibitem{Aaij:2013aln}
R.~Aaij \textit{et al.} [LHCb], Differential branching fraction and angular analysis of the decay $B_s^0\to\phi\mu^{+}\mu^{-}$, JHEP \textbf{07}, 084 (2013)
[arXiv:1305.2168 [hep-ex]].

\bibitem{Aaij:2015esa}
R.~Aaij \textit{et al.} [LHCb], Angular analysis and differential branching fraction of the decay $B^0_s\to\phi\mu^+\mu^-$, JHEP \textbf{09}, 179 (2015)
[arXiv:1506.08777 [hep-ex]].

\bibitem{Aaij:2020ruw}
R.~Aaij \textit{et al.} [LHCb], Angular Analysis of the  $B^{+}\rightarrow K^{\ast+}\mu^{+}\mu^{-}$ Decay,
Phys. Rev. Lett. \textbf{126}, no.16, 161802 (2021)
[arXiv:2012.13241 [hep-ex]].


\bibitem{Aebischer:2019mlg}
J.~Aebischer, W.~Altmannshofer, D.~Guadagnoli, M.~Reboud, P.~Stangl and D.~M.~Straub, $B$-decay discrepancies after Moriond 2019, Eur. Phys. J. C \textbf{80}, no.3, 252 (2020)
[arXiv:1903.10434 [hep-ph]].

\bibitem{Altmannshofer:2021qrr}
W.~Altmannshofer and P.~Stangl, New physics in rare B decays after Moriond 2021, Eur. Phys. J. C \textbf{81}, no.10, 952 (2021)
doi:10.1140/epjc/s10052-021-09725-1
[arXiv:2103.13370 [hep-ph]].

\bibitem{Alguero:2021anc}
M.~Alguer\'o, B.~Capdevila, S.~Descotes-Genon, J.~Matias and M.~Novoa-Brunet, $\boldsymbol{b\to s\ell\ell}$ global fits after Moriond 2021 results, [arXiv:2104.08921 [hep-ph]].

\bibitem{Alguero:2019ptt}
M.~Alguer\'o, B.~Capdevila, A.~Crivellin, S.~Descotes-Genon, P.~Masjuan, J.~Matias, M.~Novoa Brunet and J.~Virto, Emerging patterns of New Physics with and without Lepton Flavour Universal contributions,
Eur. Phys. J. C \textbf{79}, no.8, 714 (2019)
[arXiv:1903.09578 [hep-ph]].

\bibitem{Geng:2021nhg}
L.~S.~Geng, B.~Grinstein, S.~J\"{a}ger, S.~Y.~Li, J.~Martin Camalich and R.~X.~Shi, Implications of new evidence for lepton-universality violation in $b \to s\ell^+ \ell^-$ decays,
Phys. Rev. D \textbf{104}, no.3, 035029 (2021)
[arXiv:2103.12738 [hep-ph]].

\bibitem{Hurth:2021nsi}
T.~Hurth, F.~Mahmoudi, D.~M.~Santos and S.~Neshatpour, More Indications for Lepton Nonuniversality in $b \to s \ell^+ \ell^-$, [arXiv:2104.10058 [hep-ph]].

\bibitem{Angelescu:2021lln}
A.~Angelescu, D.~Be\v{c}irevi\'c, D.~A.~Faroughy, F.~Jaffredo and O.~Sumensari, Single leptoquark solutions to the B-physics anomalies, Phys. Rev. D \textbf{104}, no.5, 055017 (2021)
[arXiv:2103.12504 [hep-ph]].

\bibitem{Carvunis:2021jga}
A.~Carvunis, F.~Dettori, S.~Gangal, D.~Guadagnoli and C.~Normand,  On the effective lifetime of $B_{s} \to \mu\mu\gamma$, JHEP \textbf{12}, 078 (2021)
[arXiv:2102.13390 [hep-ph]].

\bibitem{Greljo:2022jac}
A.~Greljo, J.~Salko, A.~Smolkovi\v{c} and P.~Stangl, Rare $b$ decays meet high-mass Drell-Yan, [arXiv:2212.10497 [hep-ph]].

\bibitem{Alguero:2023jeh}
M.~Alguer\'o, A.~Biswas, B.~Capdevila, S.~Descotes-Genon, J.~Matias and M.~Novoa-Brunet, To (b)e or not to (b)e: No electrons at LHCb,
[arXiv:2304.07330 [hep-ph]].


\bibitem{Lees:2012xj} 
  J.~P.~Lees {\it et al.} [BaBar Collaboration],
  Evidence for an excess of $\bar{B} \to D^{(*)} \tau^-\bar{\nu}_\tau$ decays,
  Phys.\ Rev.\ Lett.\  {\bf 109}, 101802 (2012)
  [arXiv:1205.5442 [hep-ex]].
  
\bibitem{Lees:2013uzd} 
  J.~P.~Lees {\it et al.} [BaBar Collaboration],
  Measurement of an Excess of $\bar{B} \to D^{(*)}\tau^- \bar{\nu}_\tau$ Decays and Implications for Charged Higgs Bosons,
  Phys.\ Rev.\ D {\bf 88}, no. 7, 072012 (2013)
  [arXiv:1303.0571 [hep-ex]].


\bibitem{Huschle:2015rga} 
  M.~Huschle {\it et al.} [Belle Collaboration],
  Measurement of the branching ratio of $\bar{B} \to D^{(\ast)} \tau^- \bar{\nu}_\tau$ relative to $\bar{B} \to D^{(\ast)} \ell^- \bar{\nu}_\ell$ decays with hadronic tagging at Belle,
  Phys.\ Rev.\ D {\bf 92}, no. 7, 072014 (2015)
  [arXiv:1507.03233 [hep-ex]].


\bibitem{Sato:2016svk} 
  Y.~Sato {\it et al.} [Belle Collaboration],
  Phys.\ Rev.\ D {\bf 94}, no. 7, 072007 (2016)
  [arXiv:1607.07923 [hep-ex]].
  
\bibitem{Hirose:2017vbz} 
  S.~Hirose [Belle Collaboration],
  $\bar{B} \rightarrow D^{(*)} \tau^- \bar{\nu}_\tau$ and Related Tauonic Topics at Belle,
  arXiv:1705.05100 [hep-ex].
  
\bibitem{Aaij:2015yra} 
  R.~Aaij {\it et al.} [LHCb Collaboration],
  Measurement of the ratio of branching fractions $\mathcal{B}(\bar{B}^0 \to D^{*+}\tau^{-}\bar{\nu}_{\tau})/\mathcal{B}(\bar{B}^0 \to D^{*+}\mu^{-}\bar{\nu}_{\mu})$,
  Phys.\ Rev.\ Lett.\  {\bf 115}, no. 11, 111803 (2015)
  Erratum: [Phys.\ Rev.\ Lett.\  {\bf 115}, no. 15, 159901 (2015)]
  [arXiv:1506.08614 [hep-ex]].


\bibitem{Aaij:2017deq} 
  R.~Aaij {\it et al.} [LHCb Collaboration],
  Test of Lepton Flavor Universality by the measurement of the $B^0 \to D^{*-} \tau^+ \nu_{\tau}$ branching fraction using three-prong $\tau$ decays,
  Phys.\ Rev.\ D {\bf 97}, no. 7, 072013 (2018)
  [arXiv:1711.02505 [hep-ex]].


\bibitem{Aaij:2017uff} 
  R.~Aaij {\it et al.} [LHCb Collaboration],
  Measurement of the ratio of the $B^0 \to D^{*-} \tau^+ \nu_{\tau}$ and $B^0 \to D^{*-} \mu^+ \nu_{\mu}$ branching fractions using three-prong $\tau$-lepton decays,
  Phys.\ Rev.\ Lett.\  {\bf 120}, no. 17, 171802 (2018)
  [arXiv:1708.08856 [hep-ex]].

\bibitem{Belle:2019rba}
G.~Caria \textit{et al.} [Belle Collaboration], Measurement of $\mathcal{R}(D)$ and $\mathcal{R}(D^*)$ with a Semileptonic Tagging Method, 
Phys. Rev. Lett. \textbf{124} (2020) no.16, 161803
[arXiv:1910.05864 [hep-ex]].

\bibitem{Hirose:2017dxl} 
  S.~Hirose {\it et al.} [Belle Collaboration],
  Measurement of the $\tau$ lepton polarization and $R(D^*)$ in the decay $\bar{B} \rightarrow D^* \tau^- \bar{\nu}_\tau$ with one-prong hadronic $\tau$ decays at Belle,
  Phys.\ Rev.\ D {\bf 97}, no. 1, 012004 (2018)
  [arXiv:1709.00129 [hep-ex]].


\bibitem{Hirose:2016wfn} 
  S.~Hirose {\it et al.} [Belle Collaboration],
  Measurement of the $\tau$ lepton polarization and $R(D^*)$ in the decay $\bar{B} \to D^* \tau^- \bar{\nu}_\tau$,
  Phys.\ Rev.\ Lett.\  {\bf 118}, no. 21, 211801 (2017)
  [arXiv:1612.00529 [hep-ex]].


\bibitem{Aaij:2017tyk} 
R.~Aaij {\it et al.} (LHCb Collaboration), Measurement of the ratio of branching fractions $\mathcal{B}(B_c^+\,\to\,J/\psi\tau^+\nu_\tau)$/$\mathcal{B}(B_c^+\,\to\,J/\psi\mu^+\nu_\mu)$,  Phys. Rev. Lett. \textbf{120}, 121801 (2018) \href{http://arxiv.org/abs/1711.05623}{[arXiv:1711.05623 [hep-ex]]}.

\bibitem{Abdesselam:2019wbt} 
A.~Abdesselam {\it et al.} [Belle Collaboration], Measurement of the $D^{\ast-}$ polarization in the decay $B^0 \to D^{\ast -}\tau^+\nu_{\tau}$,
  arXiv:1903.03102 [hep-ex].


\bibitem{HFLAV:2022pwe}
Y.~Amhis \textit{et al.} [HFLAV], Averages of $b$-hadron, $c$-hadron, and $\tau$-lepton properties as of 2021,
[arXiv:2206.07501 [hep-ex]].
  
\bibitem{LHCb2022}
LHCb Collaboration, First joint measurement of $R(D^\ast)$ and $R(D^0)$ at LHCb, \url{https://indico.cern.ch/event/1187939/}

\bibitem{LHCb:2023zxo}
 [LHCb], Measurement of the ratios of branching fractions $\mathcal{R}(D^{*})$ and $\mathcal{R}(D^{0})$,
[arXiv:2302.02886 [hep-ex]].

\bibitem{LHCb2023}
LHCb Collaboration, Measurement of $R(D^\ast)$ with hadronic $\tau^+$ decays at $\sqrt{s}=$ 13 TeV by the LHCb collaboration, \url{https://indico.cern.ch/event/1231797/}.

\bibitem{HFLAVsummer}
For updated results see HFLAV preliminary average of $R(D^{(\ast)})$ for Winter 2023 in \url{https://hflav-eos.web.cern.ch/hflav-eos/semi/winter23_prel/html/RDsDsstar/RDRDs.html}.

\bibitem{Harrison:2020nrv}
J.~Harrison \textit{et al.} [LATTICE-HPQCD], $R(J/\psi)$ and $B_c^- \rightarrow J/\psi \ell^-\bar{\nu}_\ell$ Lepton Flavor Universality Violating Observables from Lattice QCD,
Phys. Rev. Lett. \textbf{125}, no.22, 222003 (2020)
[arXiv:2007.06956 [hep-lat]].


\bibitem{Iguro:2022yzr}
S.~Iguro, T.~Kitahara and R.~Watanabe, Global fit to $b \to c\tau\nu$ anomalies 2022 mid-autumn,
[arXiv:2210.10751 [hep-ph]].

\bibitem{Alonso:2016oyd} 
R.~Alonso, B.~Grinstein and J.~Martin Camalich, Lifetime of $B_c^-$ Constrains Explanations for Anomalies in  $B\to D^{(*)}\tau\nu$, Phys.\ Rev.\ Lett.\  {\bf 118}, 081802 (2017). \href{http://arxiv.org/abs/1611.06676}{[arXiv:1611.06676 [hep-ph]]}   

\bibitem{Akeroyd:2017mhr}
A.~G.~Akeroyd and C.~H.~Chen, Constraint on the branching ratio of $B_c \to \tau \nu$ from LEP1 and consequences for R(D(*)) anomaly, Phys. Rev. D \textbf{96}, 075011 (2017). \href{http://arxiv.org/abs/1708.04072}{[arXiv:1708.04072 [hep-ph]]}.

\bibitem{Kamali:2018bdp}
S.~Kamali, New physics in inclusive semileptonic $B$ decays including nonperturbative corrections,
Int. J. Mod. Phys. A \textbf{34}, no.06n07, 1950036 (2019)
[arXiv:1811.07393 [hep-ph]].

\bibitem{LHCb:2022piu}
R.~Aaij \textit{et al.} [LHCb], Observation of the decay $ \Lambda_b^0\rightarrow \Lambda_c^+\tau^-\overline{\nu}_{\tau}$,
Phys. Rev. Lett. \textbf{128}, no.19, 191803 (2022)
[arXiv:2201.03497 [hep-ex]].

\bibitem{Fedele:2022iib}
M.~Fedele, M.~Blanke, A.~Crivellin, S.~Iguro, T.~Kitahara, U.~Nierste and R.~Watanabe, Impact of \ensuremath{\Lambda}b\textrightarrow{}\ensuremath{\Lambda}c\ensuremath{\tau}\ensuremath{\nu} measurement on new physics in b\textrightarrow{}cl\ensuremath{\nu} transitions,
Phys. Rev. D \textbf{107}, no.5, 055005 (2023)
[arXiv:2211.14172 [hep-ph]].

\bibitem{Garcia-Duque:2022tti}
C.~H.~Garc\'\i{}a-Duque, J.~M.~Cabarcas, J.~H.~Mu\~noz, N.~Quintero and E.~Rojas, Singlet vector leptoquark model facing recent LHCb and BABAR measurements,''
Nucl. Phys. B \textbf{988}, 116115 (2023)
[arXiv:2209.04753 [hep-ph]].

\bibitem{Bernlochner:2018bfn}
F.~U.~Bernlochner, Z.~Ligeti, D.~J.~Robinson and W.~L.~Sutcliffe, Precise predictions for $\Lambda_b \to \Lambda_c$ semileptonic decays, Phys. Rev. D \textbf{99}, no.5, 055008 (2019)
[arXiv:1812.07593 [hep-ph]].


\bibitem{Calibbi:2015kma}
L.~Calibbi, A.~Crivellin and T.~Ota, Effective Field Theory Approach to $b\to s\ell\ell^{(\prime)}$, $B\to  K^{(*)}\nu\overline{\nu}$ and $B\to D^{(*)}\tau\nu$ with Third  Generation Couplings, Phys. Rev. Lett. \textbf{115}, 181801 (2015)
[arXiv:1506.02661 [hep-ph]].

\bibitem{Greljo:2015mma}
A.~Greljo, G.~Isidori and D.~Marzocca, On the breaking of Lepton Flavor Universality in B decays, 
JHEP \textbf{07}, 142 (2015)
[arXiv:1506.01705 [hep-ph]].

\bibitem{Bhattacharya:2014wla}
B.~Bhattacharya, A.~Datta, D.~London and S.~Shivashankara, Simultaneous Explanation of the $R_K$ and $R(D^{(*)})$ Puzzles, Phys. Lett. B \textbf{742}, 370-374 (2015)
[arXiv:1412.7164 [hep-ph]].

\bibitem{Faroughy:2016osc} 
D.~A.~Faroughy, A.~Greljo and J.~F.~Kamenik, Confronting lepton flavor universality violation in B decays with high-$p_T$ tau lepton searches at LHC, Phys.\ Lett.\ B {\bf 764}, 126 (2017). \href{http://arxiv.org/abs/1609.07138}{[arXiv:1609.07138 [hep-ph]]} 

\bibitem{Buttazzo:2017ixm}
D.~Buttazzo, A.~Greljo, G.~Isidori and D.~Marzocca, B-physics anomalies: a guide to combined explanations,
JHEP \textbf{11}, 044 (2017)
[arXiv:1706.07808 [hep-ph]].

\bibitem{Bhattacharya:2016mcc}
B.~Bhattacharya, A.~Datta, J.~P.~Gu\'evin, D.~London and R.~Watanabe, Simultaneous explanation of the $R_K$ and $R_{D^{(*)}}$ puzzles: a model analysis, JHEP \textbf{01}, 015 (2017)
[arXiv:1609.09078 [hep-ph]].

\bibitem{Kumar:2018kmr}
J.~Kumar, D.~London and R.~Watanabe, Combined Explanations of the $b \to s \mu^+ \mu^-$ and $b \to c \tau^- {\bar\nu}$ Anomalies: a General Model Analysis,
Phys. Rev. D \textbf{99}, no.1, 015007 (2019)
[arXiv:1806.07403 [hep-ph]].

\bibitem{Guadagnoli:2018ojc}
D.~Guadagnoli, M.~Reboud and O.~Sumensari, A gauged horizontal $SU(2)$ symmetry and $R_{K^{(\ast)}}$, JHEP \textbf{11}, 163 (2018)
[arXiv:1807.03285 [hep-ph]].

\bibitem{Boucenna:2016wpr}
S.~M.~Boucenna, A.~Celis, J.~Fuentes-Martin, A.~Vicente and J.~Virto, Non-abelian gauge extensions for B-decay anomalies, Phys. Lett. B \textbf{760}, 214-219 (2016)
[arXiv:1604.03088 [hep-ph]].

\bibitem{Boucenna:2016qad}
S.~M.~Boucenna, A.~Celis, J.~Fuentes-Martin, A.~Vicente and J.~Virto, Phenomenology of an $SU(2) \times SU(2) \times U(1)$ model with lepton-flavour non-universality,
JHEP \textbf{12}, 059 (2016)
[arXiv:1608.01349 [hep-ph]].

\bibitem{Capdevila:2020rrl}
B.~Capdevila, A.~Crivellin, C.~A.~Manzari and M.~Montull, Explaining $b\to s\ell^+\ell^-$ and the Cabibbo angle anomaly with a vector triplet, Phys. Rev. D \textbf{103}, no.1, 015032 (2021)
[arXiv:2005.13542 [hep-ph]].


\bibitem{Gomez:2019xfw}
J.~D.~G\'omez, N.~Quintero and E.~Rojas, Charged current $b \to c \tau \bar{\nu}_\tau$ anomalies in a general $W^\prime$ boson scenario, Phys. Rev. D \textbf{100}, no.9, 093003 (2019)
[arXiv:1907.08357 [hep-ph]].

\bibitem{Datta:2017aue}
A.~Datta, S.~Kamali, S.~Meinel and A.~Rashed, Phenomenology of $ {\Lambda}_b\to {\Lambda}_c\tau {\overline{\nu}}_{\tau } $ using lattice QCD calculations, JHEP \textbf{08}, 131 (2017)
[arXiv:1702.02243 [hep-ph]].


\bibitem{Belle-II:2018jsg}
E.~Kou \textit{et al.} [Belle-II], The Belle II Physics Book, 
PTEP \textbf{2019}, no.12, 123C01 (2019)
[erratum: PTEP \textbf{2020}, no.2, 029201 (2020)]
[arXiv:1808.10567 [hep-ex]].

\cite{Glattauer:2015teq}
\bibitem{Glattauer:2015teq}
R.~Glattauer \textit{et al.} [Belle], Measurement of the decay $B\to D\ell\nu_\ell$ in fully reconstructed events and determination of the Cabibbo-Kobayashi-Maskawa matrix element $|V_{cb}|$,
Phys. Rev. D \textbf{93}, no.3, 032006 (2016)
[arXiv:1510.03657 [hep-ex]].

\bibitem{Belle:2017rcc}
A.~Abdesselam \textit{et al.} [Belle], Precise determination of the CKM matrix element $\left| V_{cb}\right|$ with $\bar B^0 \to D^{*\,+} \, \ell^- \, \bar \nu_\ell$ decays with hadronic tagging at Belle,
[arXiv:1702.01521 [hep-ex]].

\bibitem{Becirevic:2020rzi}
D.~Be\v{c}irevi\'c, F.~Jaffredo, A.~Pe\~nuelas and O.~Sumensari, New Physics effects in leptonic and semileptonic decays, JHEP \textbf{05}, 175 (2021)
[arXiv:2012.09872 [hep-ph]].

\bibitem{Bobeth:2021lya}
C.~Bobeth, M.~Bordone, N.~Gubernari, M.~Jung and D.~van Dyk, Lepton-flavour non-universality of ${\bar{B}}\rightarrow D^*\ell {{\bar{\nu }}}$ angular distributions in and beyond the Standard Model,
Eur. Phys. J. C \textbf{81}, no.11, 984 (2021)
[arXiv:2104.02094 [hep-ph]].

\bibitem{UTfit:2022hsi}
M.~Bona \textit{et al.} [UTfit],
``New UTfit Analysis of the Unitarity Triangle in the Cabibbo-Kobayashi-Maskawa scheme,''
[arXiv:2212.03894 [hep-ph]].

\bibitem{PDG2020}
R.~L.~Workman \textit{et al.} [Particle Data Group], Review of Particle Physics,
PTEP \textbf{2022}, 083C01 (2022)

\bibitem{Belle:2019iji}
M.~T.~Prim \textit{et al.} [Belle], Search for $B^+ \to \mu^+\, \nu_\mu$ and $B^+ \to \mu^+\, N$ with inclusive tagging, Phys. Rev. D \textbf{101}, no.3, 032007 (2020)
[arXiv:1911.03186 [hep-ex]].


\bibitem{Garcia-Duque:2021qmg}
C.~H.~Garc\'\i{}a-Duque, J.~H.~Mu\~noz, N.~Quintero and E.~Rojas, Extra gauge bosons and lepton flavor universality violation in $\Upsilon$ and $B$ meson decays,
Phys. Rev. D \textbf{103}, no.7, 073003 (2021)
[arXiv:2103.00344 [hep-ph]].

\bibitem{Aloni:2017eny}
D.~Aloni, A.~Efrati, Y.~Grossman and Y.~Nir, $\Upsilon$ and $\psi$ leptonic decays as probes of solutions to the $R_D^{(*)}$ puzzle, JHEP \textbf{06}, 019 (2017)
[arXiv:1702.07356 [hep-ph]].

\bibitem{Descotes-Genon:2021uez}
S.~Descotes-Genon, S.~Fajfer, J.~F.~Kamenik and M.~Novoa-Brunet, Testing lepton flavor universality in $\Upsilon (4S)$ decays, Phys. Rev. D \textbf{103}, no.11, 113009 (2021)
[arXiv:2104.06842 [hep-ph]].

\bibitem{delAmoSanchez:2010bt}
P.~del Amo Sanchez \textit{et al.} [BaBar], Test of lepton universality in $\Upsilon(1S)$ decays at BaBar,
Phys. Rev. Lett. \textbf{104}, 191801 (2010)
[arXiv:1002.4358 [hep-ex]].

\bibitem{Besson:2006gj}
D.~Besson \textit{et al.} [CLEO], First Observation of $\Upsilon(3S) \to \tau^+ \tau^-$ and Tests of Lepton Universality in Upsilon Decays, Phys. Rev. Lett. \textbf{98}, 052002 (2007)
[arXiv:hep-ex/0607019 [hep-ex]].

\bibitem{Lees:2020kom}
J.~P.~Lees \textit{et al.} [BaBar], Precision measurement of the ${\cal B}(\Upsilon(3S)\to\tau^+\tau^-)/{\cal B}(\Upsilon(3S)\to\mu^+\mu^-)$ ratio, Phys. Rev. Lett. \textbf{125}, 241801 (2020)
[arXiv:2005.01230 [hep-ex]].

\bibitem{Belle:2022cce}
S.~Patra \textit{et al.} [Belle], Search for charged lepton flavor violating decays of $\Upsilon(1S)$,
JHEP \textbf{05}, 095 (2022)
[arXiv:2201.09620 [hep-ex]].


\bibitem{DiLuzio:2019jyq}
L.~Di Luzio, M.~Kirk, A.~Lenz and T.~Rauh, $\Delta M_s$ theory precision confronts flavour anomalies,
JHEP \textbf{12}, 009 (2019)
[arXiv:1909.11087 [hep-ph]].

\bibitem{DiLuzio:2017fdq}
L.~Di Luzio, M.~Kirk and A.~Lenz, Updated $B_s$-mixing constraints on new physics models for $b\to s\ell^+\ell^-$ anomalies, Phys. Rev. D \textbf{97}, no.9, 095035 (2018)
[arXiv:1712.06572 [hep-ph]].


\bibitem{Alok:2021pdh}
A.~K.~Alok, N.~R.~S.~Chundawat and D.~Kumar, Impact of $b \rightarrow s \ell \ell $ anomalies on rare charm decays in non-universal $Z'$ models, Eur. Phys. J. C \textbf{82}, no.1, 30 (2022)
[arXiv:2110.12451 [hep-ph]].


\bibitem{Altmannshofer:2014pba}
W.~Altmannshofer, S.~Gori, M.~Pospelov and I.~Yavin, Neutrino Trident Production: A Powerful Probe of New Physics with Neutrino Beams, Phys. Rev. Lett. \textbf{113}, 091801 (2014)
[arXiv:1406.2332 [hep-ph]].


\bibitem{Aaij:2020mqb}
R.~Aaij \textit{et al.} [LHCb], Search for the lepton flavour violating decay $B^+ \rightarrow K^+ \mu^- \tau^+$ using $B_{s2}^{*0}$ decays, JHEP \textbf{06}, 129 (2020)
[arXiv:2003.04352 [hep-ex]].

\bibitem{LHCb:2022wrs}
[LHCb], Search for the lepton-flavour violating decays $B^0 \to K^{*0} \tau^\pm \mu^\mp$, 
[arXiv:2209.09846 [hep-ex]].

\bibitem{Parrott:2022zte}
W.~G.~Parrott \textit{et al.} [HPQCD], Standard Model predictions for B\textrightarrow{}K\ensuremath{\ell}+\ensuremath{\ell}-, B\textrightarrow{}K\ensuremath{\ell}1-\ensuremath{\ell}2+ and B\textrightarrow{}K\ensuremath{\nu}\ensuremath{\nu}\textasciimacron{} using form factors from Nf=2+1+1 lattice QCD,
Phys. Rev. D \textbf{107}, no.1, 014511 (2023)
[arXiv:2207.13371 [hep-ph]].


\bibitem{Aaij:2019okb}
R.~Aaij \textit{et al.} [LHCb], Search for the lepton-flavour-violating decays $B^{0}_{s}\to\tau^{\pm}\mu^{\mp}$ and $B^{0}\to\tau^{\pm}\mu^{\mp}$, Phys. Rev. Lett. \textbf{123}, no.21, 211801 (2019)
[arXiv:1905.06614 [hep-ex]].

\bibitem{Bause:2020auq}
R.~Bause, H.~Gisbert, M.~Golz and G.~Hiller, Lepton universality and lepton flavor conservation tests with dineutrino modes, Eur. Phys. J. C \textbf{82}, no.2, 164 (2022)
[arXiv:2007.05001 [hep-ph]].

\bibitem{Bause:2021cna}
R.~Bause, H.~Gisbert, M.~Golz and G.~Hiller, Interplay of dineutrino modes with semileptonic rare B-decays, 
JHEP \textbf{12}, 061 (2021)
[arXiv:2109.01675 [hep-ph]].

\bibitem{Browder:2021hbl}
T.~E.~Browder, N.~G.~Deshpande, R.~Mandal and R.~Sinha, Impact of $B \to K \nu \bar{\nu}$ measurements on beyond the Standard Model theories, Phys. Rev. D \textbf{104}, no.5, 053007 (2021)
[arXiv:2107.01080 [hep-ph]].

\bibitem{He:2021yoz}
X.~G.~He and G.~Valencia, $R^\nu_{K^{(\ast)}}$ and non-standard neutrino interactions,
Phys. Lett. B \textbf{821}, 136607 (2021)
[arXiv:2108.05033 [hep-ph]].

\bibitem{Buras:2014fpa}
A.~J.~Buras, J.~Girrbach-Noe, C.~Niehoff and D.~M.~Straub, $ B\to {K}^{\left(\ast \right)}\nu \overline{\nu} $ decays in the Standard Model and beyond, JHEP \textbf{02}, 184 (2015)
[arXiv:1409.4557 [hep-ph]].

\bibitem{Grygier:2017tzo}
J.~Grygier \textit{et al.} [Belle], Search for $\boldsymbol{B\to h\nu\bar{\nu}}$ decays with semileptonic tagging at Belle, Phys. Rev. D \textbf{96}, no.9, 091101 (2017)
[arXiv:1702.03224 [hep-ex]].

\bibitem{Belle-II:2021rof}
F.~Abudin\'en \textit{et al.} [Belle-II], Search for $B^+ \to K^+\nu\bar{\nu}$ Decays Using an Inclusive Tagging Method at Belle II, Phys. Rev. Lett. \textbf{127}, no.18, 181802 (2021)
[arXiv:2104.12624 [hep-ex]].

\bibitem{Dattola:2021cmw}
F.~Dattola [Belle-II], Search for $B^{+} \to K^{+} \nu \bar \nu$ decays with an inclusive tagging method at the Belle II experiment, [arXiv:2105.05754 [hep-ex]].

\bibitem{Aaij:2017xqt}
R.~Aaij \textit{et al.} [LHCb], Search for the decays $B_s^0\to\tau^+\tau^-$ and $B^0\to\tau^+\tau^-$,
Phys. Rev. Lett. \textbf{118}, no.25, 251802 (2017)
[arXiv:1703.02508 [hep-ex]].

\bibitem{Bobeth:2013uxa}
C.~Bobeth, M.~Gorbahn, T.~Hermann, M.~Misiak, E.~Stamou and M.~Steinhauser, $B_{s,d} \to l^+ l^-$ in the Standard Model with Reduced Theoretical Uncertainty, Phys. Rev. Lett. \textbf{112}, 101801 (2014)
[arXiv:1311.0903 [hep-ph]].

\bibitem{Albrecht:2017odf}
J.~Albrecht, F.~Bernlochner, M.~Kenzie, S.~Reichert, D.~Straub and A.~Tully, Future prospects for exploring present day anomalies in flavour physics measurements with Belle II and LHCb, [arXiv:1709.10308 [hep-ph]].

\bibitem{Cornella:2019hct}
C.~Cornella, J.~Fuentes-Martin and G.~Isidori, Revisiting the vector leptoquark explanation of the B-physics anomalies, JHEP \textbf{07}, 168 (2019)
[arXiv:1903.11517 [hep-ph]].

\bibitem{Pich:2013lsa}
A.~Pich, Precision Tau Physics, Prog. Part. Nucl. Phys. \textbf{75}, 41-85 (2014)
[arXiv:1310.7922 [hep-ph]].

\bibitem{Belle:2023ziz}
N.~Tsuzuki \textit{et al.} [Belle], Search for lepton-flavor-violating $\tau$ decays into a lepton and a vector meson using the full Belle data sample,
[arXiv:2301.03768 [hep-ex]].

\bibitem{Langacker:2008yv}
P.~Langacker, The Physics of Heavy $Z^\prime$ Gauge Bosons,
Rev. Mod. Phys. \textbf{81}, 1199-1228 (2009)
[arXiv:0801.1345 [hep-ph]].

\bibitem{ATLAS:2019erb}
G.~Aad \textit{et al.} [ATLAS], Search for high-mass dilepton resonances using 139 fb$^{-1}$ of $pp$ collision data collected at $\sqrt{s}=$13 TeV with the ATLAS detector, Phys. Lett. B \textbf{796}, 68-87 (2019)
[arXiv:1903.06248 [hep-ex]].


\bibitem{CMS:2019tbu}
 [CMS], Search for a narrow resonance in high-mass dilepton final states in proton-proton collisions using 140$~\mathrm{fb}^{-1}$ of data at $\sqrt{s}=13~\mathrm{TeV}$,
CMS-PAS-EXO-19-019.


\bibitem{ATLAS:2017eiz}
M.~Aaboud \textit{et al.} [ATLAS], Search for additional heavy neutral Higgs and gauge bosons in the ditau final state produced in 36 fb$^{-1}$ of pp collisions at $ \sqrt{s}=13 $ TeV with the ATLAS detector,
JHEP \textbf{01} (2018), 055
[arXiv:1709.07242 [hep-ex]].

\bibitem{ATLAS:2019lsy}
G.~Aad \textit{et al.} [ATLAS], Search for a heavy charged boson in events with a charged lepton and missing transverse momentum from $pp$ collisions at $\sqrt{s} = 13$ TeV with the ATLAS detector,
Phys. Rev. D \textbf{100}, no.5, 052013 (2019)
[arXiv:1906.05609 [hep-ex]].

\bibitem{ATLASmonotau}
G.~Aad \textit{et al.} [ATLAS], Search for high-mass resonances in final states with a tau lepton and missing transverse momentum with the ATLAS detector,
ATLAS-CONF-2021-025.

\bibitem{CMS:2022ncp}
 [CMS], Search for new physics in the $\tau$ lepton plus missing transverse momentum final state in proton-proton collisions at $\sqrt s $ = 13 TeV,
[arXiv:2212.12604 [hep-ex]].

\bibitem{Erler:2011ud}
J.~Erler, P.~Langacker, S.~Munir and E.~Rojas, $Z^\prime$ Bosons at Colliders: a Bayesian Viewpoint,
JHEP \textbf{11}, 076 (2011)
[arXiv:1103.2659 [hep-ph]].


\bibitem{Salazar:2015gxa}
C.~Salazar, R.~H.~Benavides, W.~A.~Ponce and E.~Rojas, LHC Constraints on 3-3-1 Models,
JHEP \textbf{07}, 096 (2015)
[arXiv:1503.03519 [hep-ph]].

\bibitem{Benavides:2018fzm}
R.~H.~Benavides, L.~Mu\~noz, W.~A.~Ponce, O.~Rodr\'\i{}guez and E.~Rojas, Electroweak couplings and LHC constraints on alternative Z' models in $E_6$,
Int. J. Mod. Phys. A \textbf{33}, no.35, 1850206 (2018)
[arXiv:1801.10595 [hep-ph]].


\end{thebibliography}
\end{document}